\documentclass[a4paper,twoside]{article}

\usepackage{epsfig}
\usepackage{subcaption}
\usepackage{calc}
\usepackage{amssymb}
\usepackage{amstext}
\usepackage{amsmath}
\usepackage{amsthm}
\usepackage{multicol}
\usepackage{pslatex}
\usepackage{natbib}
\usepackage{algorithm2e}
\usepackage[bottom]{footmisc}
\usepackage{SCITEPRESS}     % Please add other packages that you may need BEFORE the SCITEPRESS.sty package.

\usepackage{comment}
\usepackage{enumitem}

\begin{document}

\title{A Preliminary Investigation into Theory-Practice Barriers in Sino-New Zealand Undergraduate Computing Education}

\author{\authorname{Fei Dai\sup{1}, Anthony Robins\sup{2}, Zhihao Peng\sup{3}, Wanni Huang\sup{1},  Chiu-Pih Tan\sup{1} and Tianzhen Chen\sup{3}}
\affiliation{\sup{1}School of Computing, Eastern Institute of Technology, Napier, New Zealand}
\affiliation{\sup{2}School of Computing, University of Otago, Dunedin, New Zealand}
\affiliation{\sup{3}EIT Data Science and Communication College, Zhejiang Yuexiu University, Shaoxing, Zhejiang, China}
\email{tdai@eit.ac.nz}
}


\abstract{This paper investigates the barriers hindering the effective transition from theoretical knowledge to practical application in a Sino-New Zealand double-degree undergraduate computing programme. In this unique educational setting, students study at a campus in China but complete both Chinese and New Zealand courses taught jointly by lecturers from both countries. Many enrolled students come from non-technical backgrounds, compounding the challenges they face when bridging theoretical concepts and hands-on computing tasks. Through a questionnaire administered to these students, we identify critical obstacles such as insufficient foundational knowledge, language barriers, cultural and pedagogical differences, and difficulties adapting to distinct educational systems. Our analysis reveals that these barriers significantly affect students’ academic performance, engagement, and skill development. Additionally, we examine how factors such as student interest, motivation, and career aspirations relate to these challenges. Based on the findings, we propose targeted interventions, including specialized bridging courses, enhanced language support, refined teaching methods, and improved resource allocation. These recommendations aim to foster a more supportive learning environment that helps students overcome specific theory-practice gaps. Ultimately, this study’s insights contribute to improving educational practices in Sino-New Zealand cooperative programmes and other cross-cultural computing education contexts.}

\onecolumn \maketitle \normalsize \setcounter{footnote}{0} \vfill

\section{\uppercase{Introduction}}
\label{sec:introduction}
Imagine a student enrolled in a cooperative double-degree programme hosted on a Chinese campus, where the curriculum integrates both Chinese and New Zealand courses. In their first year, students primarily focus on their literature and arts major. Starting in their second year, they transition to studying computing alongside their major courses. Lecturers from both countries contribute to the program, offering distinct pedagogical approaches, educational philosophies, and linguistic perspectives. This unique arrangement blends a literature-based foundation with a computing major, equipping students with interdisciplinary skills while exposing them to diverse academic systems. Successfully addressing the transition from theoretical concepts to practical computing skills is critical to ensuring students achieve both academic success and professional competency. This transition is not only essential for mastering complex computing concepts but also for developing the practical skills necessary for professional careers in a rapidly evolving technological landscape.

Bridging the gap between theory and practice has become increasingly critical in computing education, particularly in programs that incorporate both cultural and educational diversity. For this study, 'technology' refers to tools and systems such as Internet of Things (IoT) devices, automation systems, and the technical skills necessary for programming, data analysis, and problem-solving. Computing education demands mastering complex theoretical concepts through hands-on experimentation and application. However, the bicultural and bilingual nature of the Sino-New Zealand program introduces additional challenges, such as language barriers and divergent teaching styles, that complicate this transition.

To address these challenges, this research investigates the barriers faced by students in Electronics and IoT, Automation and Embedded Systems courses within the Sino-New Zealand cooperative program. By examining how these barriers impact educational quality and student outcomes, the study seeks to provide actionable, evidence-based solutions for enhancing learning and teaching practices. The insights gained will benefit not only this program but also similar cross-cultural and interdisciplinary educational initiatives.

Our research question is: \textit{"What barriers hinder the effective transition from theoretical knowledge to practical skills in computing education within the Sino-New Zealand double-degree programme at our University?"}

The remainder of this paper is structured as follows: Section~\ref{sec:related_work} reviews existing research and identifies gaps relevant to this study. Section~\ref{sec:research_method} details the research methodology, including survey design, data analysis techniques, and study limitations. Section~\ref{sec:results} presents the results, focusing on the learning challenges and barriers identified, students' experiences and resources, and their interests and career aspirations. Section~\ref{sec:discussion} discusses the identified barriers and proposes solutions for overcoming these barriers. Finally, Section~\ref{sec:conclusion} concludes the study and suggests directions for future research.

\section{\uppercase{Related Work}}\label{sec:related_work}
Research on bridging theory-practice gaps in computing education spans three key areas: applying theoretical knowledge to practice, identifying theory-practice barriers, and developing methods to bridge these gaps.

The first area focuses on applying knowledge from theory to practice, encompassing studies on undergraduate computing courses~\citep{eckerdal2024learning, king2021combining, nascimento2019does}, teacher's practices~\citep{king2021combining, randi2007theory, bouzguenda2013enablers, zyad2016integrating, tedre2022grand, yilmaz2016transition, catete2020aligning}, and secondary school education~\citep{hayes2023inclusive, samarasekara2024framework}. These studies highlight the importance of bridging theoretical concepts with practical skills, emphasizing hands-on experiences to reinforce learning outcomes in computing curricula. For example, \citep{eckerdal2024learning} demonstrates how project-based learning significantly improves students' ability to translate abstract concepts into practical solutions, while \citep{hayes2023inclusive} explores inclusive approaches in secondary schools that foster early computational thinking.

The second area examines the barriers that impede effective learning and application. Challenges include those faced by women in computer science education~\citep{scragg1998study, bock2013women, kordaki2020identifying}, broader student learning obstacles~\citep{pappas2017assessing, schulte2007attitudes}, and pedagogical issues from the teacher's perspective~\citep{belland2009using, aflalo2014invisible, gretter2019equitable, zyad2016integrating}. Additionally, barriers in coding bootcamps~\citep{thayer2017barriers, thayer2020practical} and high school education~\citep{wang2017diversity, samarasekara2022barriers} further highlight structural and systemic challenges. For instance, \citep{thayer2017barriers} analyze how inadequate preparation and lack of mentorship hinder the success of bootcamp graduates, while \citep{schulte2007attitudes} address the role of student attitudes in learning efficacy.

The third area explores solutions to bridge theory-practice gaps, such as computing education projects~\citep{gelonch2017teaching, martin2004addressing, giacaman2018bridging}, curriculum revisions~\citep{young2003bridging, janse2020developing}, and AI-driven tools~\citep{murtaza2024theory}. For example, \citep{giacaman2018bridging} propose a novel project-based learning framework that integrates real-world scenarios, while \citep{murtaza2024theory} discuss the application of generative AI in personalizing computing education. These approaches highlight innovative ways to adapt teaching strategies and foster experiential learning.

While existing studies offer valuable insights, they rarely capture the unique dynamics of cross-cultural, double-degree programs like the Sino-New Zealand arrangement. Students in this program must navigate significant language differences, distinct educational systems, and rigorous technical requirements, all while adapting to a computing major from a non-technical background. This context presents a set of barriers that are not fully captured by studies focusing on more uniform learning environments.

\section{\uppercase{Research Method}}\label{sec:research_method}

\subsection{Survey Design}

To investigate the learning challenges and barriers faced by students in courses that integrate both theoretical and practical tasks, we designed a structured questionnaire targeting students in the Sino-New Zealand educational programme. The questionnaire consisted of 20 multiple-choice questions and 2 open-ended questions, organized into four main sections:
(1) \textit{Demographic Information}, 
(2) \textit{Learning Challenges and Barriers}, (3) \textit{Students’ Experiences and Resources} and (4) \textit{Final Remarks}. 

Most questions employed a 5-point Likert scale ranging from "Strongly Disagree" to "Strongly Agree," with an option to skip questions to ensure participant comfort. Open-ended questions were included to capture nuanced feedback and suggestions beyond the scope of the predefined questions.

Before distribution, a pilot test was conducted with six participants from the target population to refine the questionnaire. Feedback from the pilot led to clarifying ambiguous terms, incorporating concrete examples, and streamlining the survey to minimize completion time without sacrificing data quality.

\subsection{Data Collection}

Participants were drawn from one major within the 2021 cohort of the Sino-New Zealand double-degree programme. The majority of participants had a background in literature and arts, reflecting limited exposure to science courses in high school. Data were collected using \textit{Jinshuju}~\citep{jinshuju}, a Chinese online survey platform comparable to SurveyMonkey or Qualtrics. Out of 83 initial submissions, 76 valid responses were analyzed after excluding one incomplete submission and six pilot test responses.

The survey was administered in Chinese to accommodate participants' language preferences. The first author later translated responses into English to facilitate analysis. A complete translated version of the survey is provided in the Appendix.

\subsection{Data Analysis}

The collected data were analyzed using both quantitative and qualitative methods:

\textbf{Quantitative Analysis:} Multiple-choice responses were exported to Excel and analyzed using Python to generate visualizations such as bar charts. This analysis revealed patterns and trends, identifying prevalent barriers and their impact on students.

\textbf{Qualitative Analysis:} Open-ended responses were reviewed for typographical errors and irrelevant content before coding. An iterative coding process was used to categorize responses into themes, focusing on common ideas and suggestions. These themes were quantified to highlight the most frequently mentioned challenges and solutions.

To ensure reliability, the initial analysis conducted by the first author was independently reviewed by other authors. Discrepancies were discussed and resolved through consensus. This collaborative approach enhanced the robustness of the findings by combining quantitative trends with in-depth qualitative insights.

\subsection{Study Limitations}

This study has several limitations that may affect the generalizability of its findings:

\textbf{Participant Demographics:} All participants were from a single major within the Sino-New Zealand programme and predominantly had literature and arts backgrounds. This specific demographic focus may limit the applicability of the findings to students from other disciplines or educational contexts.

\textbf{Survey Scope:} The questionnaire concentrated on student learning challenges, barriers, and experiences, without exploring broader perspectives such as faculty insights or institutional factors, which could provide additional context.

\textbf{Translation Bias:} Responses were translated into English by the first author, which could introduce bias despite careful attention to accuracy. Future studies could involve independent translators to mitigate this risk.

\textbf{Sample Size and Diversity:} The relatively small sample size and lack of demographic diversity limit the ability to generalize findings. Including participants from a broader range of academic backgrounds and disciplines in future research would enhance the scope and applicability of the study.

%\textbf{Depth of Qualitative Analysis:} While the thematic analysis identified major themes, it did not explore less common responses in-depth, potentially oversimplifying the complexity of student experiences.

Future research should address these limitations by incorporating diverse participant demographics, broader survey scopes, and additional qualitative methods such as interviews. These steps would provide deeper insights and enhance the reliability and generalizability of the findings.

\section{\uppercase{Results}}\label{sec:results}
\subsection{Learning Challenges and Barriers}
The first part of the questionnaire explored the difficulty in understanding theoretical concepts, challenges in applying theoretical knowledge to practical tasks, barriers in the theory-to-practice transition, and their impacts. We aimed to identify key barriers affecting education quality and uncover improvement areas.

\subsubsection{Difficulty in understanding theoretical concepts}
\begin{figure}[t!]
    \centering
    \vspace{-0mm}
    \includegraphics[width=\columnwidth]{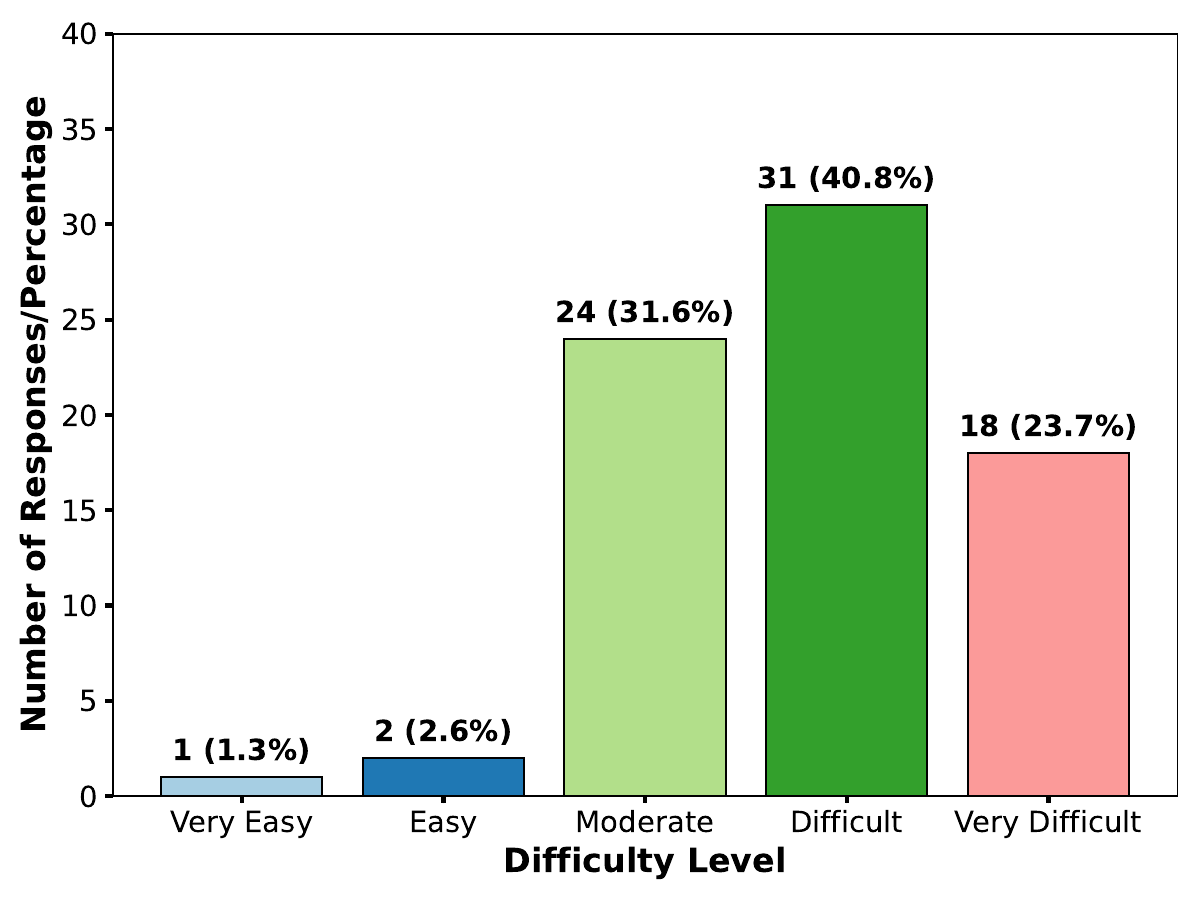}
    \caption{Difficulty Level in Understanding Theoretical Concepts}
    \label{fig:difficulty}
    \vspace{-0mm}
\end{figure}

All 76 participants rated the difficulty of understanding theoretical concepts in their computing courses. A significant majority (96.1\%) found these concepts challenging (see Fig.~\ref{fig:difficulty}). 
This indicates that many students struggle with theoretical aspects, suggesting a need for interventions to make theoretical material more accessible. Potential solutions include enhanced teaching methods and additional tutorials.

\subsubsection{Challenges in applying theoretical knowledge to practical tasks}
\begin{figure}[t!]
    \centering
    \vspace{-0mm}
    \includegraphics[width=\columnwidth]{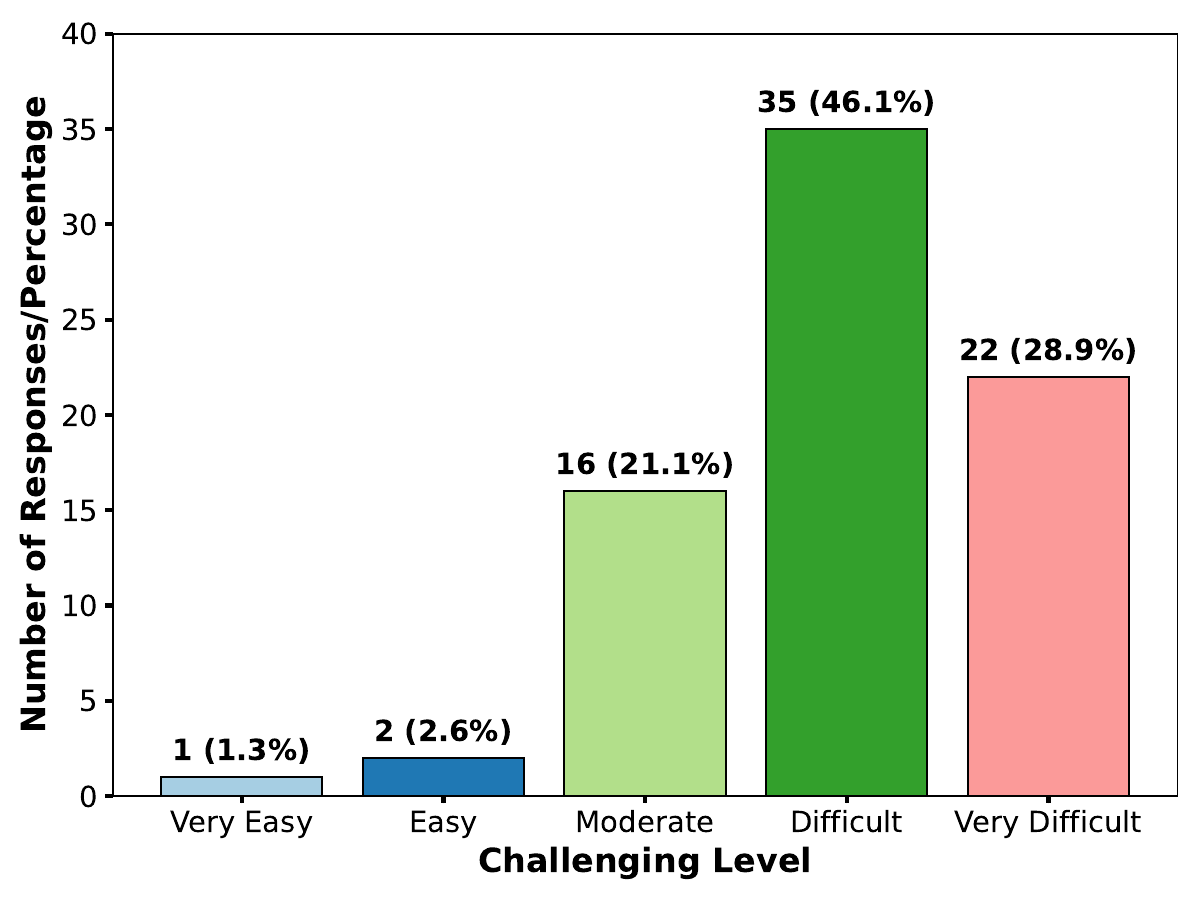}
    \caption{Challenging Level in Applying Theoretical Knowledge to Practical Tasks}
    \label{fig:Challenging}
    \vspace{-0mm}
\end{figure}

Participants were also asked about the challenging level of applying theoretical knowledge to practical tasks in their computing courses. Most students (96.1\%) reported challenges in this area (see Fig.~\ref{fig:Challenging}). This underscores a significant issue in the practical application of theoretical concepts within computing courses. The findings suggest a need for curricular adjustments or enhanced instructional strategies, such as integrating more practical exercises and providing tailored support to help students bridge the theory-practice gap.

\subsubsection{Three barriers in the transition from theory to practice}\label{sec:barriers}
\begin{figure*}[t!]
  \centering
  \vspace{-0mm}
  \begin{subfigure}[b]{0.48\textwidth}
    \centering
    \includegraphics[width=\textwidth]{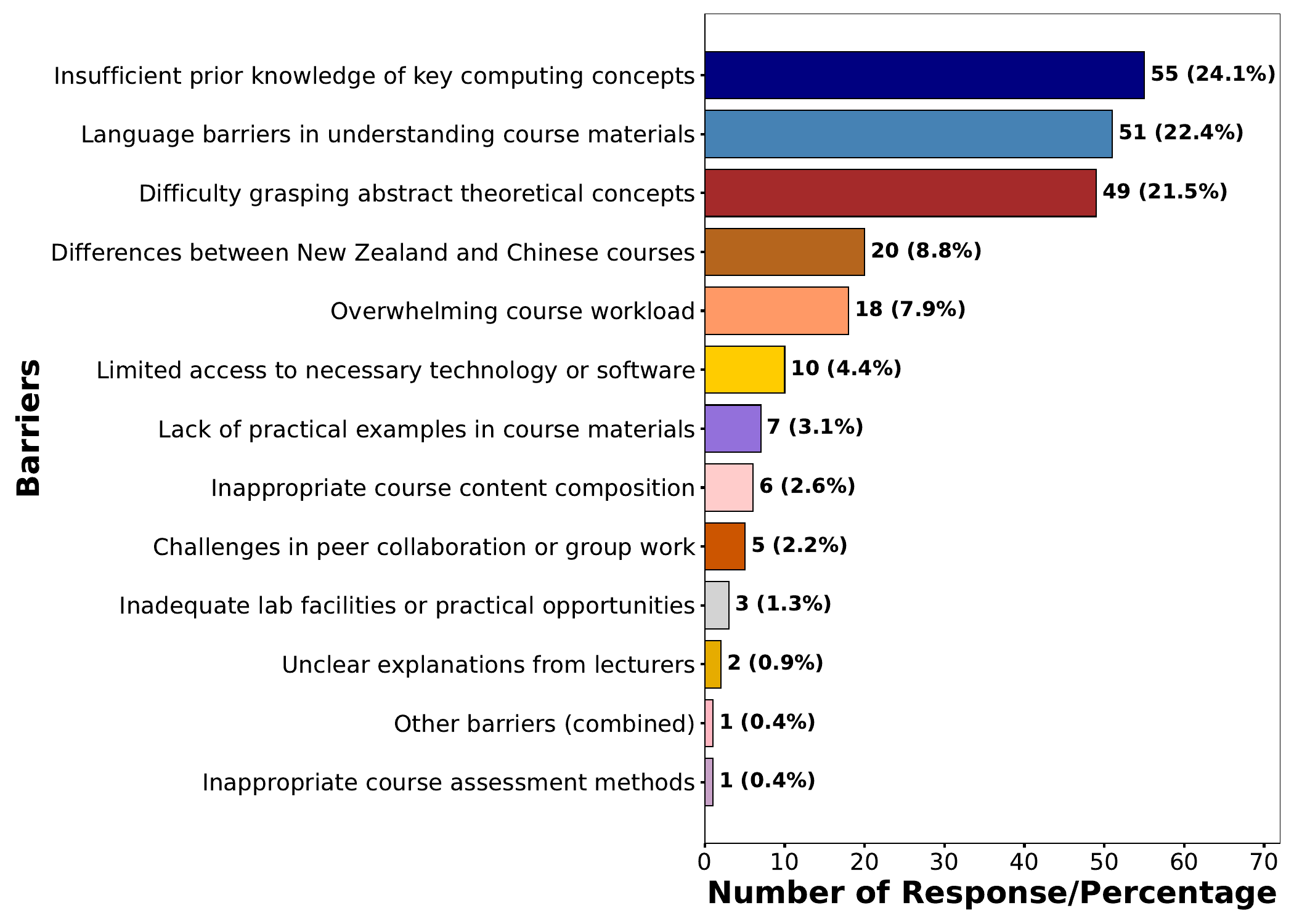}
    \caption{}
  \end{subfigure}\hfill
  \begin{subfigure}[b]{0.48\textwidth}
    \centering
    \includegraphics[width=\textwidth]{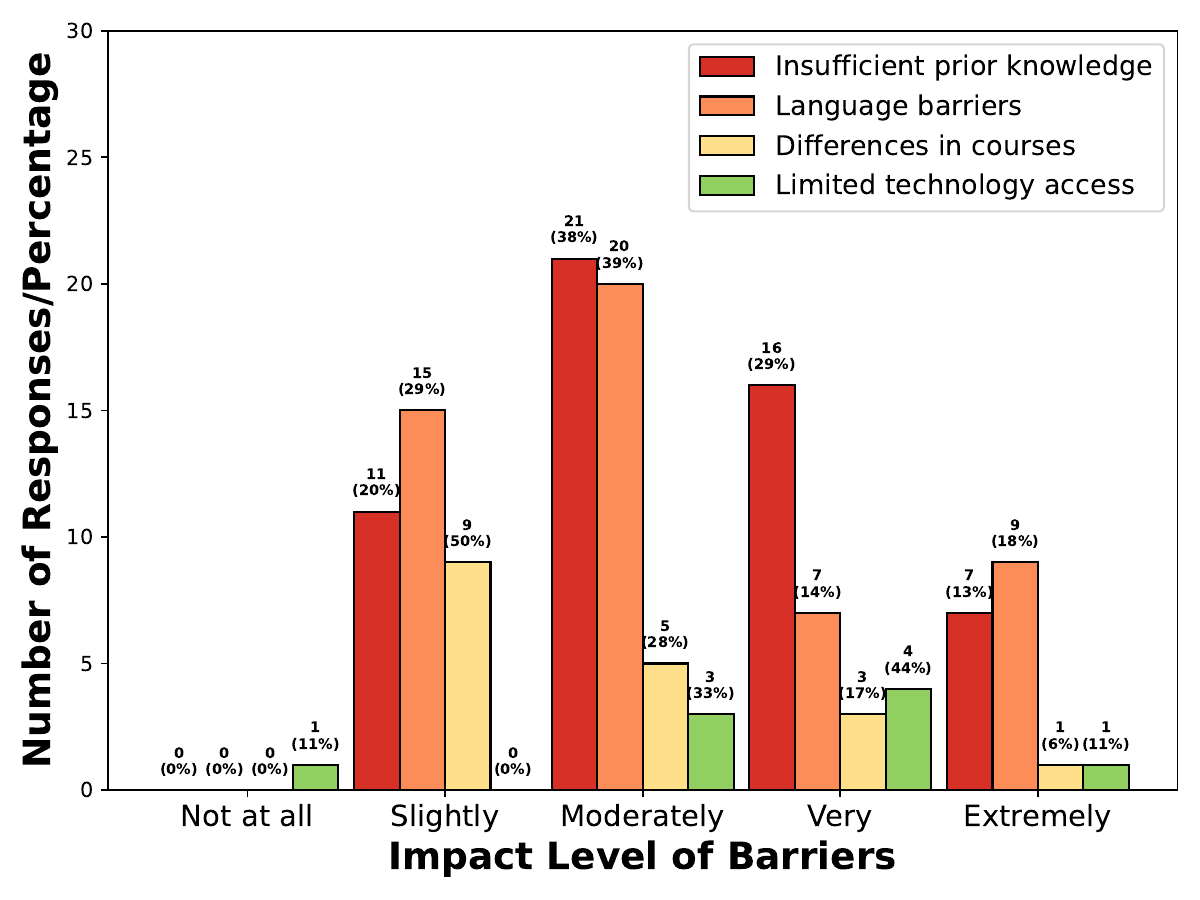}
    \caption{}
  \end{subfigure}
  \vspace{-0mm}
  \caption{(a) Barriers Distribution in Theory-to-practice Transition and (b) Impact Distribution of Barriers}
  \label{fig:barriers}
\end{figure*}

To understand the barriers students face, participants were asked to choose three barriers they believe have the worst impact on their computing courses. Eleven barrier options were provided, along with three open-ended entries for additional barriers. All 76 participants responded to this question. The top three barriers identified were:
\textbf{(1) Insufficient prior knowledge of key computing concepts};
\textbf{(2) Language barriers to understanding course materials};
\textbf{(3) Difficulty grasping abstract theoretical concepts} (see Fig.~\ref{fig:barriers} (a)). Other notable barriers included differences between New Zealand and Chinese courses, overwhelming course workload, and limited access to necessary technology or software. These findings highlight significant challenges in transitioning from theory to practice, indicating the need for targeted interventions such as foundational knowledge support, language assistance programmes, and strategies to clarify theoretical concepts.

\begin{comment}
    \begin{figure}[!hpbt]
    \centering
    \vspace{-2mm}
    \includegraphics[width=8.5cm]{barriers_chart.pdf}
    \caption{Barriers distribution in theory-to-practice transition by the participants}
    \label{fig:barriers}
    \vspace{-2mm}
\end{figure}
\end{comment}

\subsubsection{Impact of identified barriers}
Participants were asked to quantify the impact of the identified barriers on their studies. From Fig.~\ref{fig:barriers} (b), we know that insufficient prior knowledge and language barriers were reported to have moderate to severe impacts by most respondents, with some experiencing extreme difficulties. In contrast, differences between New Zealand and Chinese courses were seen as having slight to moderate impacts, indicating a less severe overall effect. Limited access to necessary technology or software showed varied impacts, emphasizing the importance of providing adequate technological resources to support learning.

The analysis highlights that insufficient prior knowledge and language barriers are the most impactful barriers to computing education at our University. Addressing these issues through targeted interventions—such as enhancing foundational knowledge, improving language support, and ensuring access to necessary technology—can significantly enhance the educational experience and academic success of computing students.

\begin{figure*}[t!]
  \centering
  \begin{subfigure}[b]{0.48\textwidth}
    \centering
    \includegraphics[width=\textwidth]{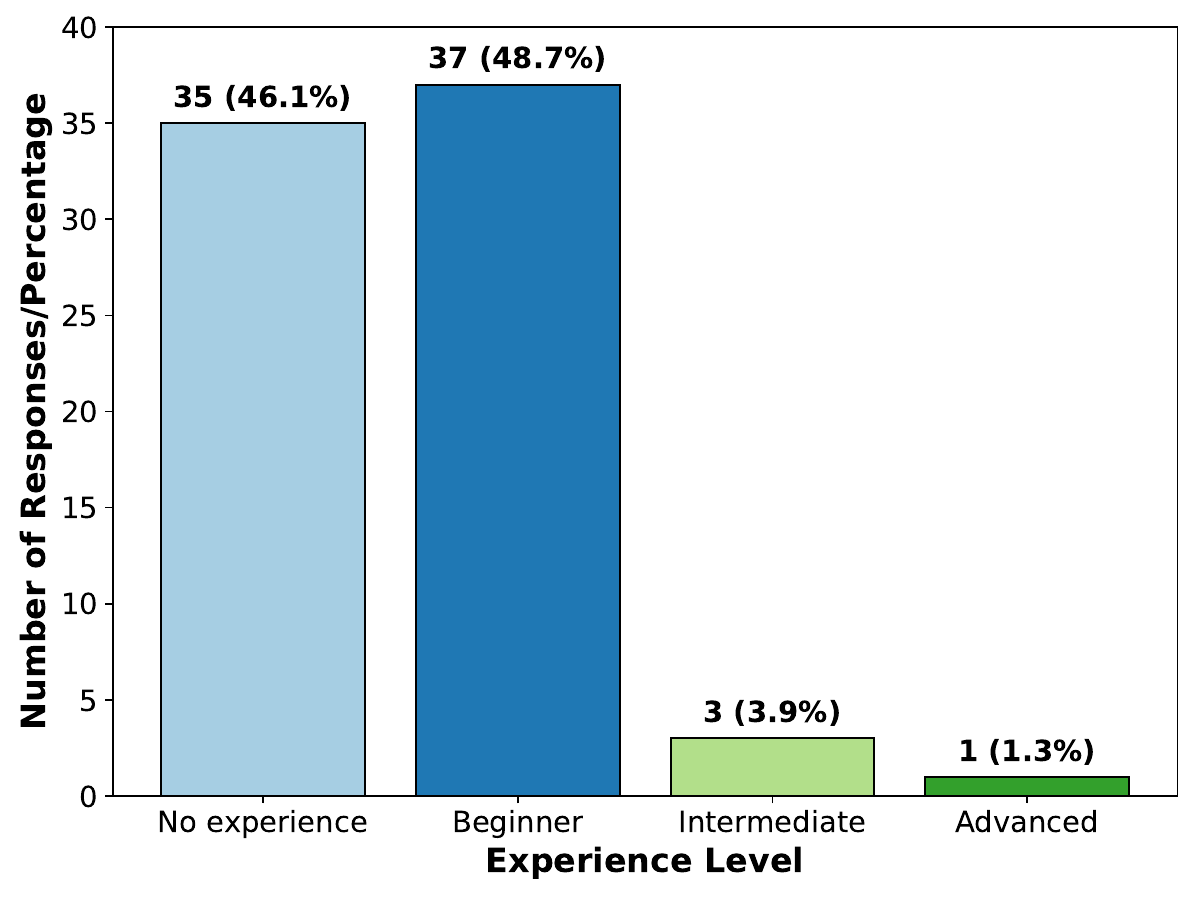}
    \vspace{-0mm}
    \caption{}
  \end{subfigure}\hfill
  \begin{subfigure}[b]{0.48\textwidth}
    \centering
    \includegraphics[width=\textwidth]{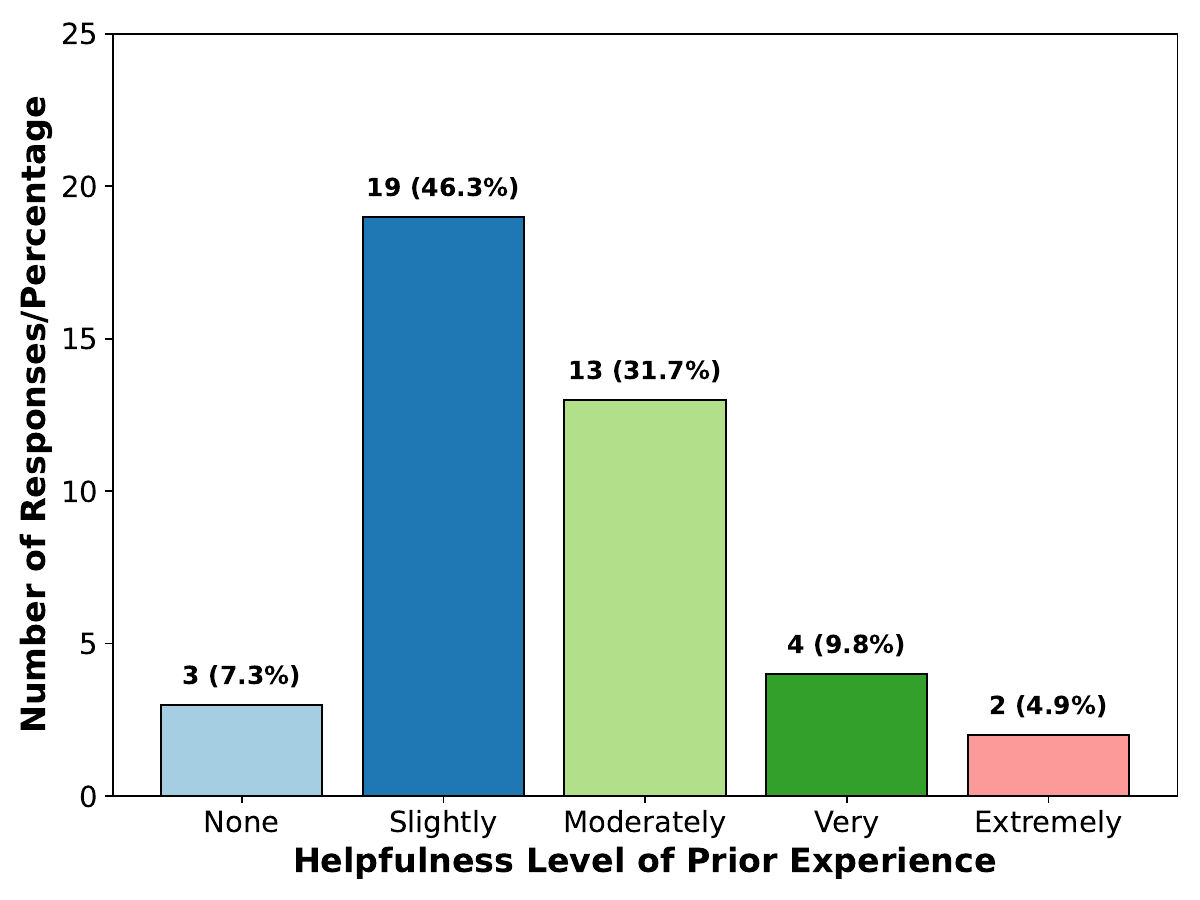}
    \vspace{-0mm}
    \caption{}
  \end{subfigure}
  \vspace{-0mm}
  \caption{(a) Distribution of Computing Experience Level and (b) Helpfulness Level of Prior Experience}
  \label{fig:experience}
\end{figure*}

\begin{comment}
    \begin{figure}[!hpbt]
    \centering
    \vspace{-2mm}
    \includegraphics[width=8.5cm]{barriers_impact.pdf}
    \caption{Distribution of Barrier Impact by the participants}
    \label{fig:impact_barrier}
    \vspace{-2mm}
\end{figure}
\end{comment}

\subsection{Students’ Experiences and Resources}
The second part of the questionnaire explored participants’ experiences of the resources provided, as well as their overall engagement. We aimed to understand how their previous educational background, the resources provided, and their engagement influenced their current learning experiences in computing courses.

\vspace{-0mm}
\subsubsection{Prior knowledge}\label{sec:prior_knowledge}
All 76 participants responded to the question about their computing experience level before entering their current courses. The vast majority (94.8\%) reported having little or no prior computing knowledge (see Fig.~\ref{fig:experience} (a)). Only a small fraction rated their prior knowledge as "Intermediate" or "Expert." When asked about the extent to which their previous computing experience helped them perform well in their courses, 53.9\% of participants responded. Of these, most found their prior experience helpful to some degree, while a small number felt it was not helpful at all (see Fig.~\ref{fig:experience} (b)).

\begin{comment}
    \begin{figure}[!hpbt]
    \centering
    \vspace{-2mm}\includegraphics[width=8.5cm]{experience.pdf}
    \caption{Distribution of computing experience level by the participants}
    \label{fig:experience}
    \vspace{-2mm}
\end{figure}

\begin{figure}[!hpbt]
    \centering
    \vspace{-2mm}\includegraphics[width=8.5cm]{help_prior.pdf}
    \caption{Helpfulness level of prior experience by the participants}
    \label{fig:help_prior}
    \vspace{-2mm}
\end{figure}
\end{comment}

These findings indicate that most students entered the computing programme with limited prior experience, which may contribute to the challenges they face in grasping and applying theoretical concepts in practical scenarios. While those with prior experience generally found it beneficial, the extent of its helpfulness varied. This suggests that previous exposure alone may not fully support students in advanced courses without additional resources and support. The results underscore the importance of providing robust introductory courses and supplementary resources to bridge the knowledge gap for students with limited backgrounds. Enhancing support systems, such as tutoring and mentoring programmes, could aid students in their transition and improve their performance in computing courses. Addressing these gaps is crucial for fostering a more inclusive and effective learning environment that accommodates students of varying backgrounds and experience levels.

\begin{figure*}[t!]
  \centering
  \begin{subfigure}[b]{0.48\textwidth}
    \centering
    \includegraphics[width=\textwidth]{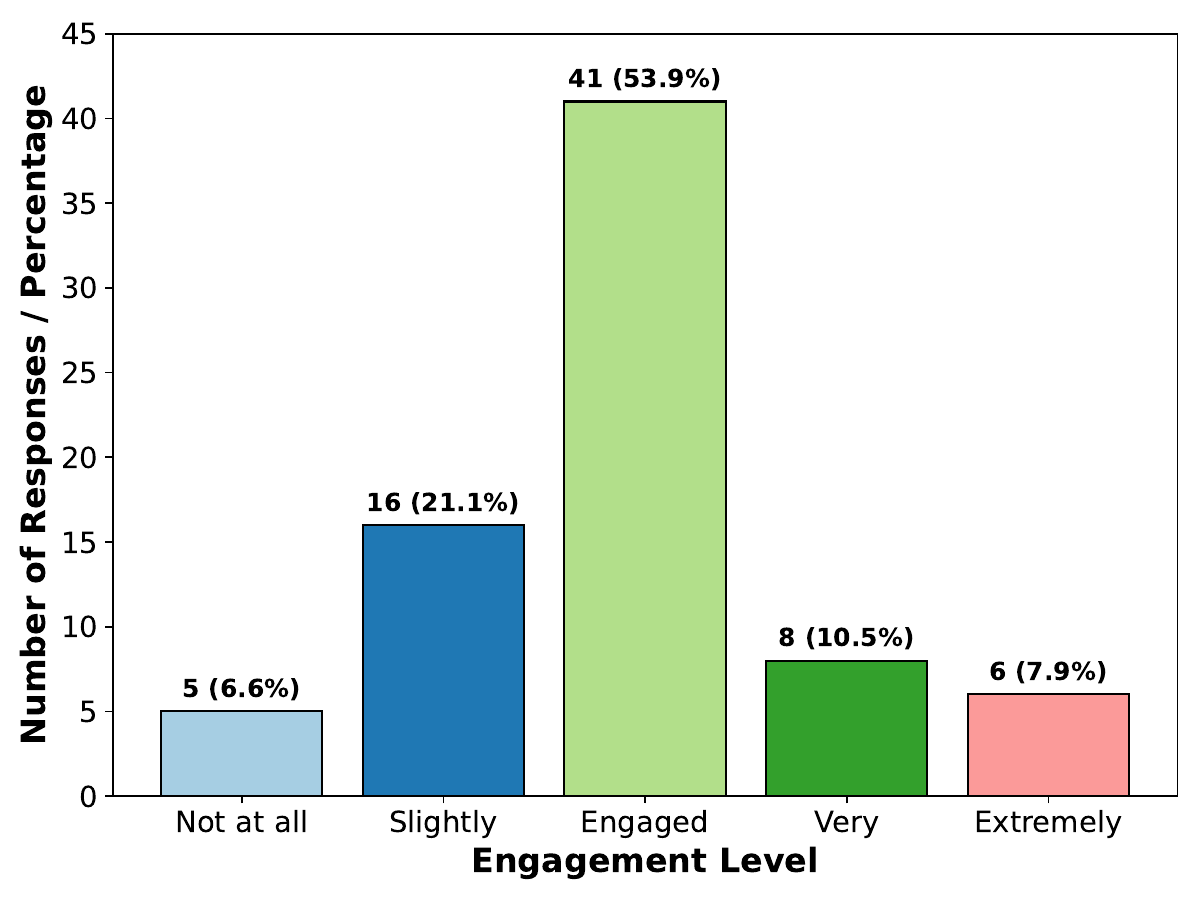}
    \vspace{-0mm}
    \caption{}
  \end{subfigure}\hfill
  \begin{subfigure}[b]{0.48\textwidth}
    \centering
    \includegraphics[width=\textwidth]{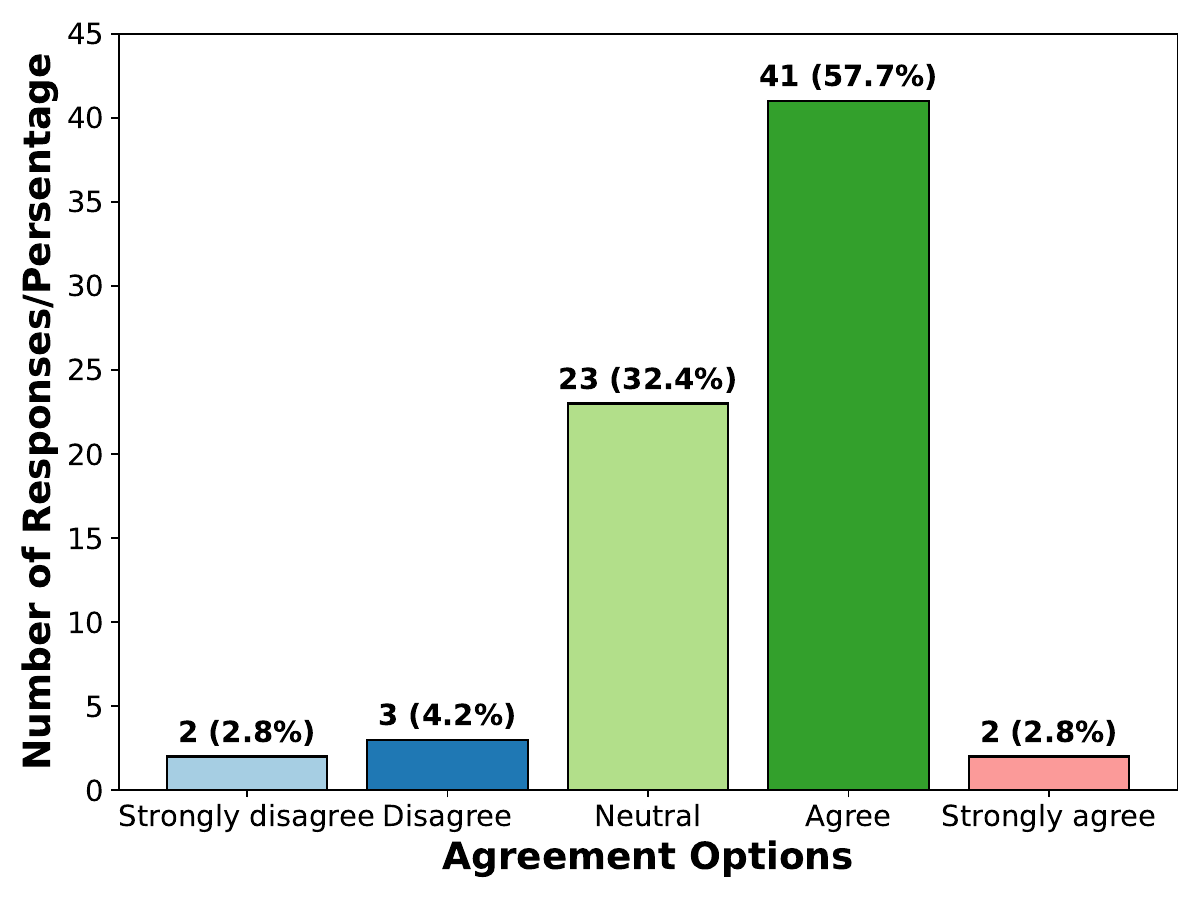}
    \vspace{-0mm}
    \caption{}
  \end{subfigure}
  \vspace{-0mm}
  \caption{(a) Engagement Level for Computing Courses and (b) Engagement Impact for Computing Courses}
  \label{fig:engagement}
\end{figure*}

\vspace{-0mm}
\subsubsection{Engagement and its impact}
All 76 participants responded to the question about their level of engagement in computing courses. The majority reported being engaged to some extent, with 93.4\% indicating some level of engagement and only 6.6\% stating they were "minimal" engaged (see Fig.~\ref{fig:engagement} (a)). When asked whether their level of engagement affects their learning and helps overcome learning barriers, most participants agreed, with 60.5\% expressing agreement and 32.4\% holding a neutral perspective (see Fig.~\ref{fig:engagement} (b)). Only a small percentage (7\%) disagreed.

\begin{comment}
    \begin{figure}[!hpbt]
    \centering
    \vspace{-2mm}
    \includegraphics[width=8.5cm]{engagement.pdf}
    \caption{Engagement level for computing course by the participants}
    \label{fig:engagement}
    \vspace{-2mm}
\end{figure}

\begin{figure}[!hpbt]
    \centering
    \vspace{-2mm}
    \includegraphics[width=8.5cm]{engagement_impact.pdf}
    \caption{Engagement impact for computing course by the participants}
    \label{fig:engagement_impact}
    \vspace{-2mm}
\end{figure}
\end{comment}

These findings suggest that most students are engaged in their computing courses and believe this enhances their learning and helps them overcome barriers. However, some neutral and disagreeing responses suggest that engagement alone may not address all the challenges students face. This underscores the importance of fostering interest through interactive teaching methods and collaborative projects, as well as providing ample support and resources to boost participation. By addressing these aspects, the university can enhance the overall educational experience for computing students.

\begin{figure}[t!]
    \centering
    \vspace{-0mm}\includegraphics[width=\columnwidth]{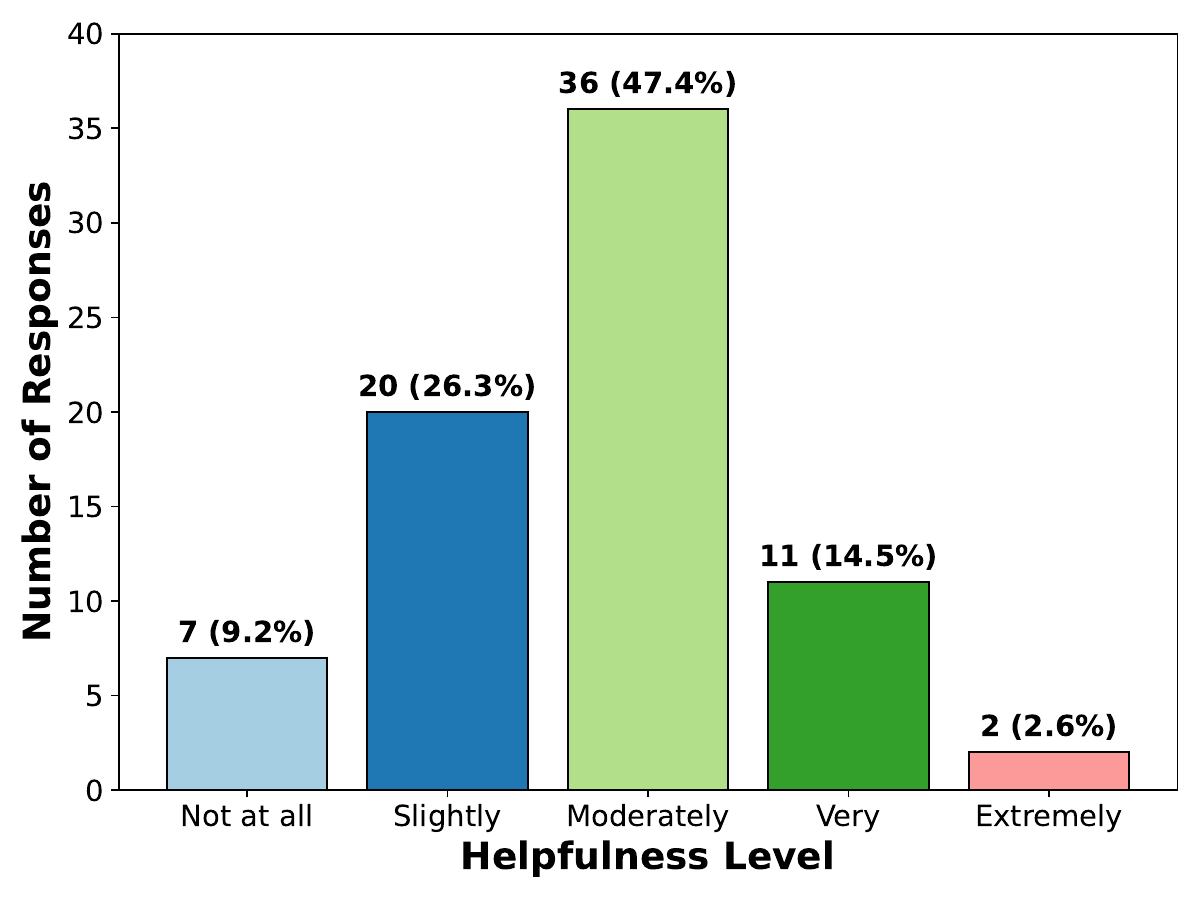}
    \caption{Helpfulness Level from Lecturer and TA}
    \label{fig:lecturer_help}
    \vspace{-0mm}
\end{figure}

\subsubsection{Support from lecturer and teaching assistant}
Participants were asked about the extent to which support from lecturers and teaching assistants (TA) helped them overcome barriers. All 76 participants responded, with 90.8\% indicating that the support was helpful to some degree, and only 9.2\% stating they gained no help (see Fig.~\ref{fig:lecturer_help}).
These results indicate that support from lecturer and TA plays a crucial role in helping students overcome barriers in their computing courses. The majority found the support at least somewhat helpful, highlighting the importance of accessible and supportive instructors. Enhancing this support, perhaps through additional training and resources from a lecturer and TA, could further assist students in navigating academic challenges.

\subsubsection{Impact of the transition from Chinese to New Zealand educational systems}
\begin{figure}[t!]
    \centering
    \vspace{-0mm}
    \includegraphics[width=\columnwidth]{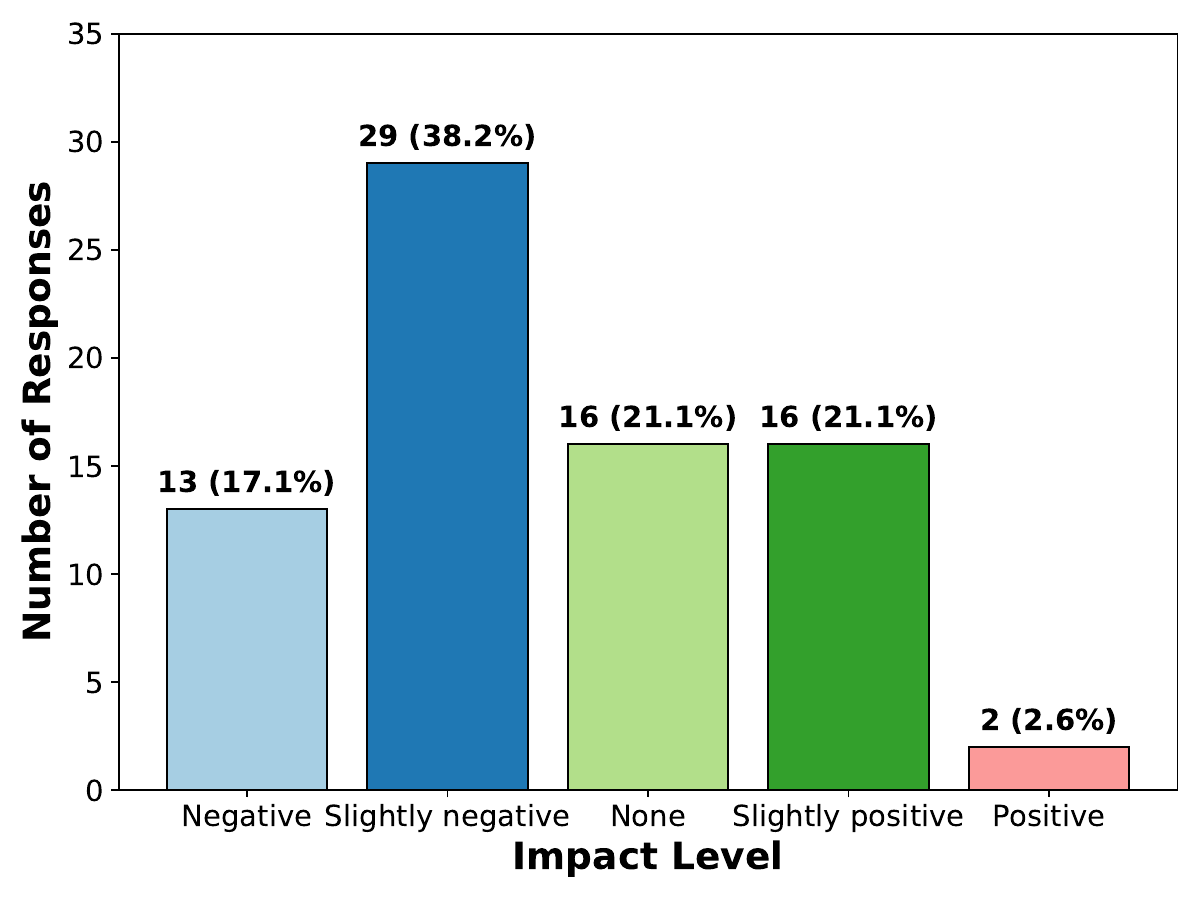}
    \caption{Transition Impact from Different Education Systems}
    \label{fig:transition_impact}
    \vspace{-0mm}
\end{figure}

All 76 participants responded to the question about how various cultural teaching methodologies, such as independent learning versus rote memorization-based approaches, influence their learning experiences and outcomes. Responses revealed three distinct viewpoints: 55.3\% perceived the transition as having a negative impact on their learning experience, 21.1\% reported no impact, and 23.7\% perceived a positive impact (see Fig.~\ref{fig:transition_impact}).
These findings suggest that while the transition from the Chinese to the New Zealand education system poses challenges for many students, a notable portion either adapt without significant effect or find the new educational environment beneficial. The data underscores the importance of providing targeted support to help students navigate differences in educational methodologies. Strategies such as orientation programmes, peer support networks, and cultural competency training for educators can facilitate smoother transitions and enhance learning experiences for international students.

\vspace{-0mm}
\subsubsection{Advice on improving courses}
In their open-ended responses, students offered several suggestions for enhancing the courses. Key themes emerged, including:

\textbf{Emphasize English Proficiency and Fundamental Knowledge:} Students (N=7) express language and technical barriers in their study. Two students advised, "We need to enhance our English and computer background, then carry out relevant practice." %(S10 and S22)

\textbf{Strengthen the explanation of technical terms/knowledge in lecture and provide more demonstrations in lab:} Students (N=6) request more explanations with examples to help them understand these abstract concepts and need more demonstrations of how to use IoT components. One mentioned: "Lecturers could focus more on the practical aspects of courses and present theory more interestingly and understandably".

\textbf{Provide in-time translation in class and after-class videos to help students better understand:} Students (N=3) request in-time translation and more videos to help them learn and understand. One stated: "Increase the range of available after-class videos and translations to help students better understand the material".

These suggestions highlight areas where the university can improve to support students' learning experiences better.
%The analysis of responses to suggestions for improvement reveals several key themes. Many students suggested increasing practical opportunities to enhance their understanding of theoretical concepts, with one student specifically mentioning, \textbf{"Increase practice opportunities."} Others felt that the difficulty level of assignments should be reduced to make them more manageable, as reflected in the advice, \textbf{"The difficulty of computer course assignments can be reduced."} Additionally, there was a call for more resources, such as after-class videos and translation services, to aid comprehension. One respondent recommended, \textbf{"Increase after-class videos and translation scope to help students better understand."} Finally, students emphasized the need for clearer explanations and simpler course content, with specific feedback including, \textbf{"Strengthen the explanation of technical terms and simplify course content,"} and \textbf{"Teachers can give examples from daily life when explaining theoretical knowledge to help everyone understand."}

\subsection{Interest and Career Aspiration}
The third part of the questionnaire explored participants’ interests and career aspirations. We aim to understand how these factors influenced their current learning experiences in computing courses.

\subsubsection{Interest in computing field}\label{sec:interest}
\begin{figure}[t!]
    \centering
    \vspace{-0mm}
    \includegraphics[width=\columnwidth]{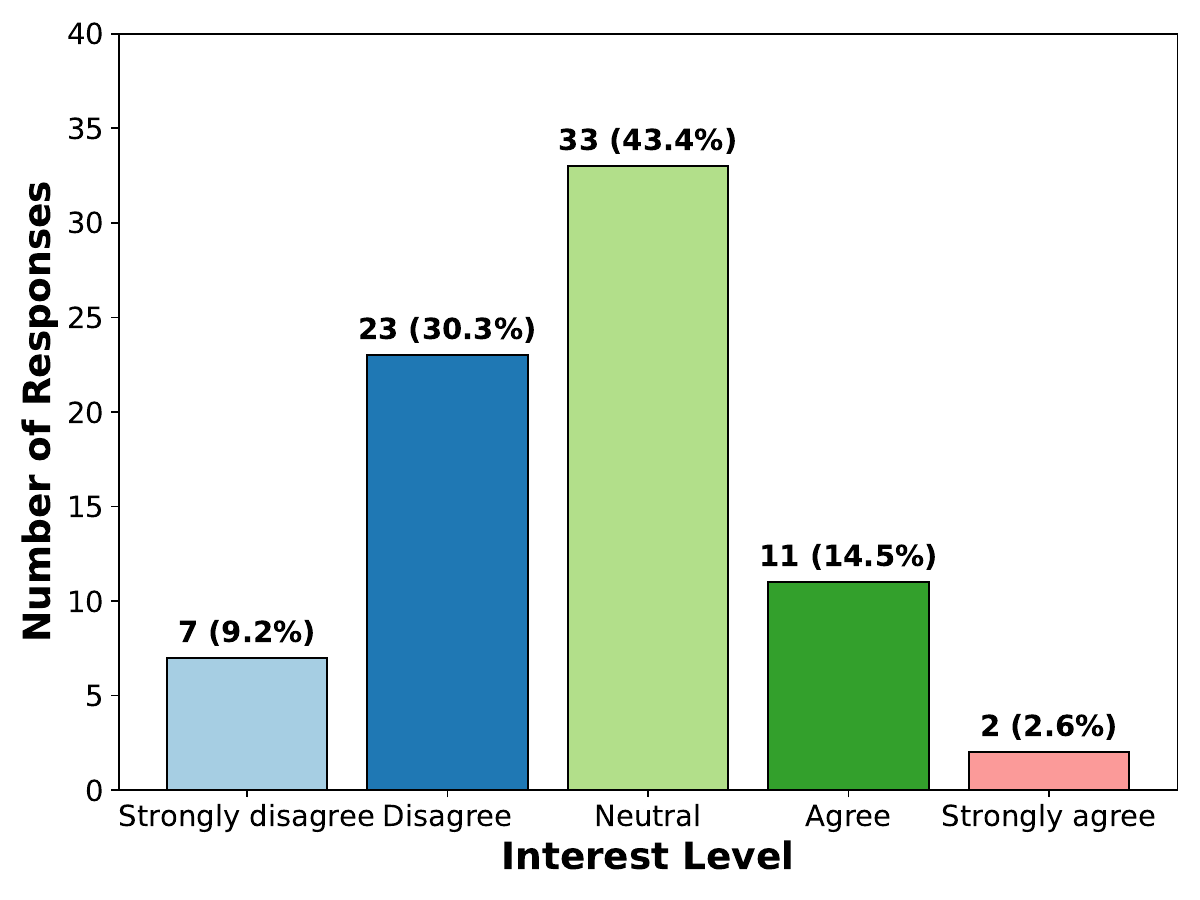}
    \caption{Interest Distribution in Computing Field}
    \label{fig:interest}
    \vspace{-0mm}
\end{figure}

Participants were asked about their interest in the computing/computer science field. As shown in Fig.~\ref{fig:interest}, a significant portion (43.4\%) remained neutral, neither agreeing nor disagreeing, while 39.5\% expressed a lack of interest in computing. Only 17.1\% indicated that they are interested in computing. These findings suggest that many students are indifferent or uninterested in computing, with only a minority expressing interest. This highlights the need for strategies to foster and increase student interest, such as showcasing diverse career opportunities, integrating engaging and hands-on projects into the curriculum, and providing mentorship to help students discover and cultivate their interest in the field.

\vspace{-0mm}
\subsubsection{Career aspiration}\label{sec:career}
\begin{figure*}[t!]
  \centering
  \begin{subfigure}[b]{0.48\textwidth}
    \centering
    \includegraphics[width=\textwidth]{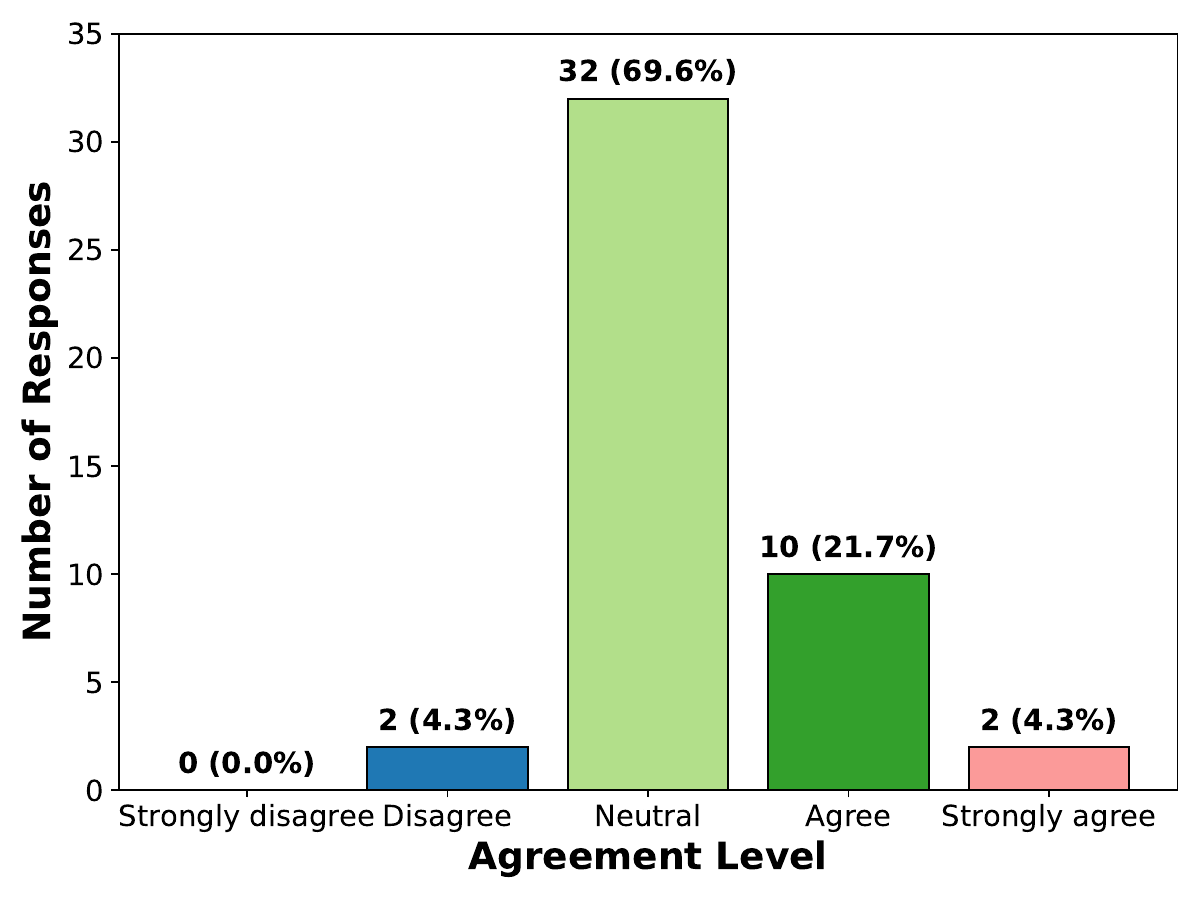}
    \vspace{-0mm}
    \caption{}
  \end{subfigure}\hfill
  \begin{subfigure}[b]{0.48\textwidth}
    \centering
    \includegraphics[width=\textwidth]{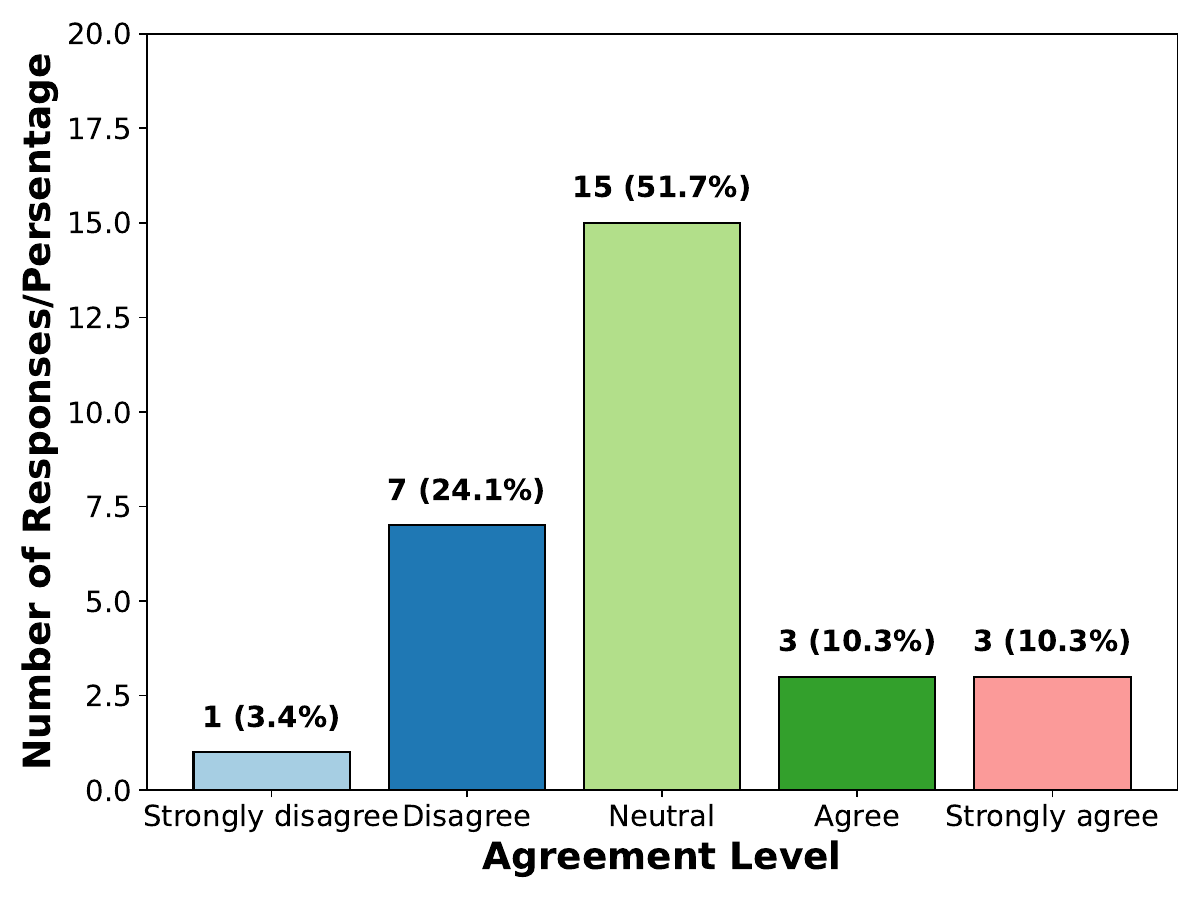}
    \vspace{-0mm}
    \caption{}
  \end{subfigure}
  \vspace{-0mm}
  \caption{(a) Alignment of Career Aspirations with Interests and (b) Influence of Parents, Relatives, and Friends on Career Aspirations}
  \label{fig:interest_career}
\end{figure*}

For students who did not disagree with the statement regarding their interest in computing, we further examined the alignment between their career aspirations and interests. As illustrated in Fig.~\ref{fig:interest_career} (a), the majority (69.6\%) neither agreed nor disagreed, indicating a neutral stance, while 26\% agreed that their career aspirations align with their interests. Only a small portion disagreed. This suggests that while some students have a clear alignment between their career goals and interests, the majority remain undecided or neutral.

For students who disagreed with the statement regarding their interest in computing, we explored whether parents, relatives, and friends influence their career aspirations. As shown in Fig.~\ref{fig:interest_career} (b), over half (51.7\%) neither agreed nor disagreed, 27.5\% disagreed, and 20.6\% agreed that their career aspirations are affected by family and friends. This suggests that while external factors influence some students in their career decisions, many remain neutral or unaffected.

\begin{comment}
\begin{figure}[!hpbt]
    \centering
    \vspace{-2mm}
    \includegraphics[width=8.5cm]{interest_career.pdf}
    \caption{Alignment of career aspirations with interests}
    \label{fig:interest_career}
    \vspace{-2mm}
\end{figure}

\begin{figure}[!hpbt]
    \centering
    \vspace{-2mm}
    \includegraphics[width=8.5cm]{carreer_parents.pdf}
    \caption{ Influence of parents, relatives, and friends on career aspirations}
    \label{fig:carreer_parents}
    \vspace{-2mm}
\end{figure}
\end{comment}

These findings highlight the importance of providing career guidance and support to help students align their career aspirations with their interests. Understanding the influence of family and friends on career decisions can aid in developing strategies to support students in making independent and informed choices.

\subsubsection{Motivation related to career}
\begin{figure*}[t!]
  \centering
  \begin{subfigure}[b]{0.48\textwidth}
    \centering
    \includegraphics[width=\textwidth]{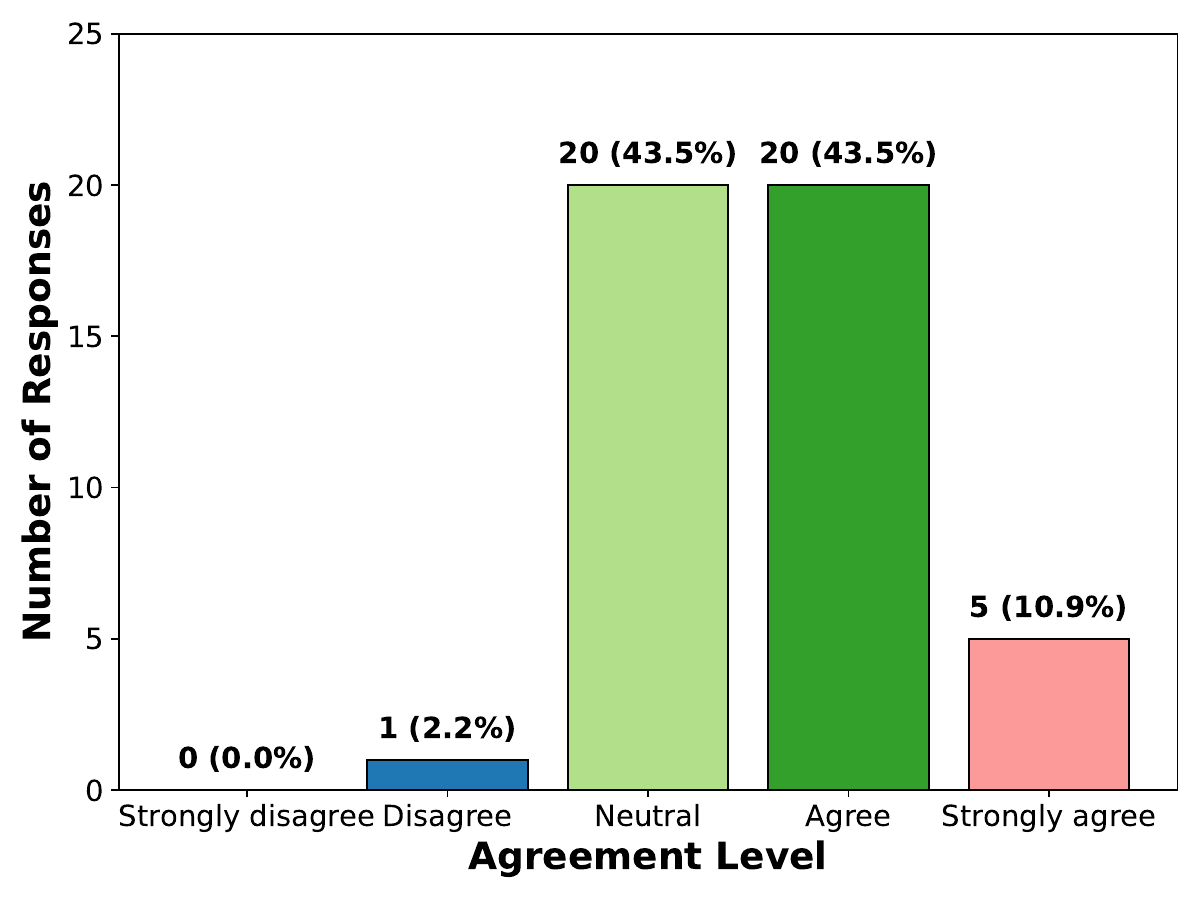}
    \vspace{-0mm}
    \caption{}
  \end{subfigure}\hfill
  \begin{subfigure}[b]{0.48\textwidth}
    \centering
    \includegraphics[width=\textwidth]{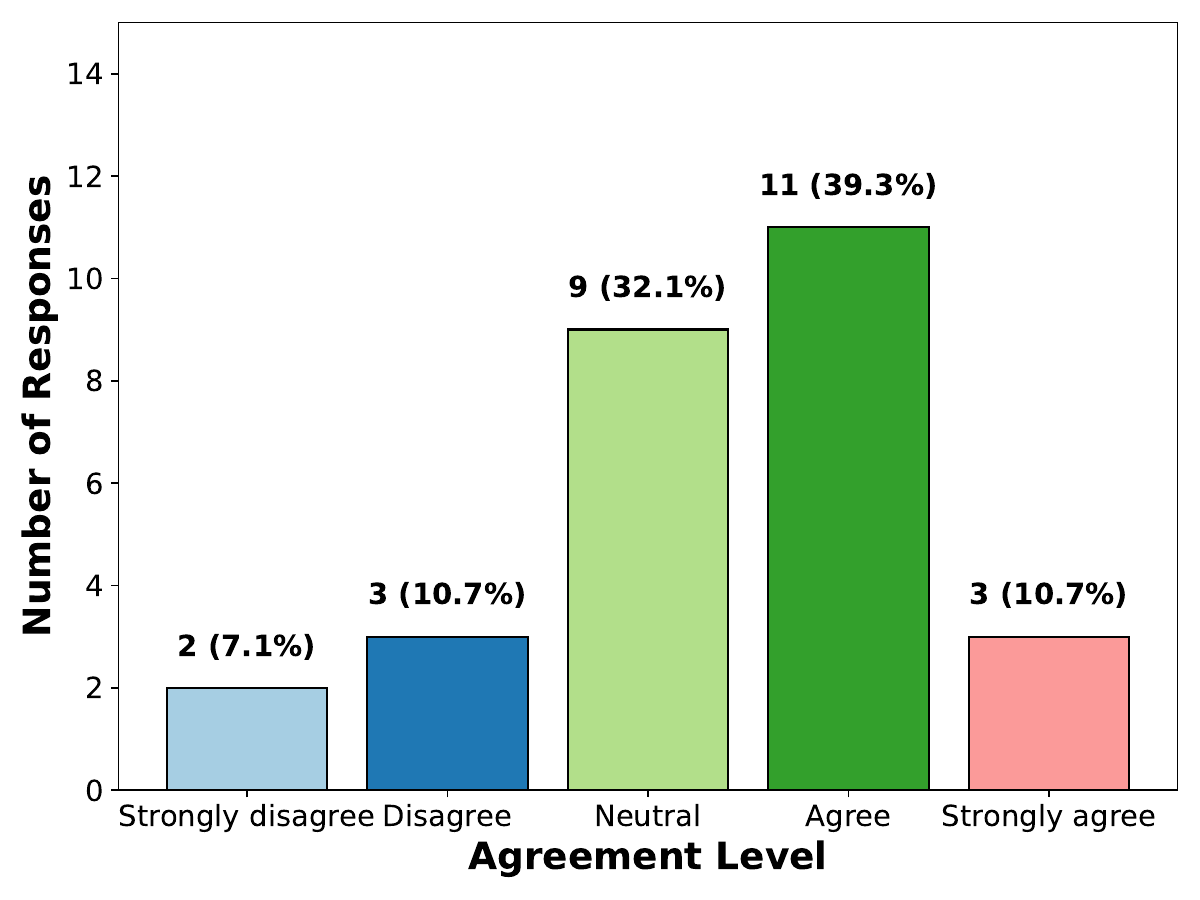}
    \vspace{-0mm}
    \caption{}
  \end{subfigure}
  \vspace{-0mm}
  \caption{(a) Motivation from Career Prospects and (b) Difficulty in Course due to Non-computing Career}
  \label{fig:motivation_career}
\end{figure*}

Participants whose career aspirations align with their interests were asked whether a career in computing motivates them and eases overcoming course barriers. As shown in Fig.~\ref{fig:motivation_career} (a), over half (54.4\%) agreed, 43.5\% were neutral, and a few disagreed. This suggests that alignment between career goals and interests is a significant motivator, though the substantial neutral response implies other factors also influence motivation.

Participants influenced by parents, relatives, or friends were asked whether pursuing a non-computing career adds difficulty to engaging with the course. Fig.~\ref{fig:motivation_career} (b) shows that half agreed, 32.1\% were neutral, and a small fraction disagreed, indicating that external influences may complicate engagement with computing courses.

Overall, alignment between career aspirations and computing interests significantly motivates students and helps them overcome academic barriers, while external influences can pose additional challenges. These findings underscore the importance of aligning career guidance with students' interests and providing support to navigate external pressures. Institutions can enhance student motivation and engagement by connecting coursework with computing career opportunities, offering mentorship, and providing resources to help students reconcile their career aspirations with their academic pursuits.

\begin{comment}
    
\begin{figure}[!hpbt]
    \centering
    \vspace{-2mm}
    \includegraphics[width=8.5cm]{motivation_career.pdf}
    \caption{ Motivation from Career Prospects}
    \label{fig:motivation_career}
    \vspace{-2mm}
\end{figure}

\begin{figure}[!hpbt]
    \centering
    \vspace{-2mm}
    \includegraphics[width=8.5cm]{no_interest.pdf}
    \caption{Difficulty in Course due to Non-computing Career}
    \label{fig:no_interest}
    \vspace{-2mm}
\end{figure}
\end{comment}

\subsubsection{Advice on study strategy or study resources for junior students}
In their open-ended responses, students offered valuable advice for junior students beginning their computing studies. Key themes emerged, emphasizing the importance of foundational skills and proactive learning approaches.

\textbf{Maintain Open Communication with Instructors:} Students (N=7) emphasized the importance of interacting with teachers. One suggested, "Learn about computer knowledge and strengthen contact with teachers." Building relationships with instructors can provide additional support, clarify doubts, and enhance overall understanding.

\textbf{Engage Attentively in Class and Practice Regularly:} Students (N=7) stressed the need to be attentive during lectures and to engage in consistent practice. One comment encapsulated this sentiment: "Listen carefully in class and practice more." Active participation and regular practice are seen as key to internalizing computing concepts.

\textbf{Focus on Learning English and Computer Knowledge:} Students (N=6) highlighted the necessity of learning English alongside foundational computer knowledge. One advised, "Focus on learning English and computer knowledge." Proficiency in English is crucial, given that much of the computing literature and resources are in this language.

These insights highlight areas where the university can enhance support for students, such as offering language assistance programmes, facilitating programming workshops, encouraging active classroom participation, and promoting open communication channels between students and faculty.

\section{\uppercase{Discussion}}
\label{sec:discussion}
\subsection{Identified Barriers}

This study sought to address the research question: \textit{"What barriers hinder the effective transition from theoretical knowledge to practical skills in computing education within the Sino-New Zealand double-degree programme at our University?"} The five key barriers identified (see Fig.~\ref{fig:barriers}) are:

\begin{enumerate}[itemsep=0pt, topsep=0pt]
\item \textbf{Insufficient prior knowledge of computing concepts}
\item \textbf{Language barriers in understanding course materials}
\item \textbf{Difficulty grasping abstract theoretical concepts}
\item \textbf{Differences between New Zealand and Chinese educational systems}
\item \textbf{Overwhelming course workload}
\end{enumerate}

Students enter the programme with limited computing experience and lack clarity about career aspirations. These factors exacerbate challenges in understanding theoretical concepts and their practical applications, making \textit{insufficient prior knowledge} and \textit{difficulty grasping abstract theoretical concepts} particularly significant barriers.

Language barriers and differences in educational systems further complicate the learning process. Students transitioning from Chinese to New Zealand courses face challenges in language proficiency and adapting to different teaching methodologies. Additionally, the overwhelming double-degree course workload places additional pressure on students with limited foundational knowledge or interest in computing.

These findings highlight the complexity of the barriers and their relevance to the research question, suggesting a need for comprehensive and interconnected strategies to address these challenges.

\subsection{Proposed Solutions}

To overcome these barriers and support students in their transition from theoretical knowledge to practical skills, a multifaceted approach is essential. Below are detailed solutions supported by evidence from educational research:

\textbf{1. Enhancing Foundational Knowledge and Language Proficiency}

\textit{Barriers Addressed}: Insufficient prior knowledge, language barriers, difficulty understanding theoretical concepts.

\begin{itemize}
\item Implement preparatory or bridging courses to strengthen foundational computing knowledge before students begin advanced courses. Initiatives like these have shown significant improvements in student preparedness \citep{wang2017diversity}.
\item Provide tutoring and mentorship programmes for students with limited prior experience. Mentorship could also address overlapping issues such as cultural adjustment or language barriers, fostering a more integrated approach to student support. Peer mentorship has been shown to enhance academic performance and social integration \citep{zyad2016integrating}.
\item Establish language assistance programmes, including technical vocabulary courses and bilingual materials. Tailored language support has been linked to reduced academic stress and improved outcomes \citep{gretter2019equitable}.
\end{itemize}

\textbf{2. Bridging Educational Systems and Promoting Cultural Support}

\textit{Barriers Addressed}: Differences in educational systems, cultural differences, language barriers.

\begin{itemize}
\item Offer orientation sessions to acclimate students to differences between New Zealand and Chinese educational systems. Transition programmes have proven effective for international students \citep{phillips2005challenging}.
\item Conduct cultural integration workshops to foster mutual understanding between students and educators. These workshops could be explicitly linked with orientation sessions to provide a cohesive approach to addressing interconnected barriers such as cultural differences, language challenges, and adaptation to different educational systems. Cultural awareness reduces misunderstandings and enhances classroom dynamics \citep{hasker2007joint}.
\item Organize activities such as \textit{English Corners} to improve language skills and cultural familiarity, promoting active learning \citep{hayes2023inclusive}.
\end{itemize}

\textbf{3. Fostering Interest in Computing and Career Guidance}

\textit{Barriers Addressed}: Lack of interest in computing, unclear career pathways.

\begin{itemize}
\item Provide career counselling services to align academic content with career goals, which motivates students \citep{martin2004addressing}.
\item Establish mentorship programmes connecting students with industry professionals to inspire and guide them \citep{resch2023using}.
\item Highlight diverse applications of computing through case studies and real-world examples, which effectively increase engagement \citep{giacaman2018bridging}.
\end{itemize}

\textbf{4. Providing Technological Resources and Managing Workload}

\textit{Barriers Addressed}: Limited access to technology, difficulty with practical tasks, overwhelming workload.

\begin{itemize}
\item Ensure access to technology and software through computer labs, software licenses, and loaner laptops. Adequate resources are critical for bridging the theory-practice gap \citep{thayer2017barriers}.
\item Conduct technical training sessions to enhance proficiency with necessary tools \citep{FLOSS}.
\item Design balanced curricula integrating theory and practice while avoiding excessive workloads \citep{aflalo2014invisible}.
\item Offer time management and study skills workshops, which have been linked to reduced stress and improved academic outcomes \citep{reimer2019computer}.
\end{itemize}

\textbf{5. Adapting Teaching Methods to Increase Engagement}

\textit{Barriers Addressed}: Difficulty understanding theoretical concepts, difficulty applying theory to practice, lack of interest in computing.

\begin{itemize}
\item Incorporate interactive teaching methods, such as group discussions, hands-on activities, and practical projects, which improve engagement \citep{catete2020aligning}.
\item Implement flipped classrooms to encourage active learning during class time \citep{eckerdal2015relating}.
\item Highlight real-world applications of computing to demonstrate relevance and foster interest \citep{young2003bridging}.
\end{itemize}

Grounding these solutions in research ensures their reliability and applicability. Together, they create a supportive learning environment addressing the diverse needs of students and enhancing academic outcomes.

\begin{comment}
\subsection{Areas for Improvement}
While the proposed solutions address immediate challenges, additional areas for improvement could enrich the educational experience and ensure long-term sustainability:

\textbf{Continuous Curriculum Development} Regularly update curricula based on student feedback and industry trends to maintain relevance and engagement.

\textbf{Faculty Development} Invest in professional development to enhance teaching skills, cultural competency, and the ability to support diverse students.

\textbf{Research and Innovation} Encourage research on effective teaching methodologies and pilot innovative educational programmes.

\textbf{Expanded Support Services} Enhance student support services, including mental health resources, academic advising, and study skill workshops.

\textbf{Collaborative Learning Environments} Foster collaboration through group projects, study groups, and student organizations to build a sense of community.

\textbf{Strengthened Industry Partnerships} Develop partnerships with industry to provide internships, real-world projects, and networking opportunities, enhancing practical skills and career readiness.

Addressing these areas supports the transition from theoretical knowledge to practical skills, ensuring the programme’s sustainability and relevance in computing education.
\end{comment}

\section{\uppercase{Conclusions and Future Work}}
\label{sec:conclusion}
This study identifies significant barriers to the effective transition from theoretical knowledge to practical skills in the Sino-New Zealand double-degree computing programme. Key challenges include insufficient foundational knowledge, language barriers, cultural and pedagogical differences, and an overwhelming course workload. Additionally, student interest, motivation, and career aspiration can further exacerbate these challenges. Addressing these issues requires an integrated, multi-faceted approach. Strategies such as preparatory bridging courses, targeted language support, and culturally responsive teaching methods are essential to creating an inclusive and effective learning environment. The proposed interventions include tailored mentorship programmes, balancing theoretical and practical content in the curriculum, and fostering student interest through real-world applications of computing. By implementing these strategies, the programme can better equip students to overcome learning barriers and achieve academic and professional success. Furthermore, these findings offer valuable insights for similar cross-cultural educational initiatives.

Future research should broaden the scope by including diverse student cohorts across various majors and educational contexts. Longitudinal studies are essential to evaluate the long-term impacts of the proposed interventions on students' academic performance and career trajectories. Comparative analyses with other international cooperative programmes could also identify universal challenges and effective solutions. Additionally, exploring the integration of innovative pedagogical approaches, such as blended learning and adaptive instruction, could provide further insights into improving educational outcomes. This continued research effort is critical for ensuring that students are well-prepared to navigate and succeed in the dynamic and evolving field of computing.

%\section*{\uppercase{Acknowledgements}}

%If any, should be placed before the references section without numbering. To do so please use the following command:
%\textit{$\backslash$section*\{ACKNOWLEDGEMENTS\}}

\bibliographystyle{apalike}
{\small
\bibliography{CSEDU-initial}}

\begin{thebibliography}{}

\bibitem[Aflalo, 2014]{aflalo2014invisible}
Aflalo, E. (2014).
\newblock The invisible barrier to integrating computer technology in education.
\newblock {\em Journal of education and learning}, 3(2):120--134.

\bibitem[Belland, 2009]{belland2009using}
Belland, B.~R. (2009).
\newblock Using the theory of habitus to move beyond the study of barriers to technology integration.
\newblock {\em Computers \& education}, 52(2):353--364.

\bibitem[Bock et~al., 2013]{bock2013women}
Bock, S.~J., Taylor, L.~J., Phillips, Z.~E., and Sun, W. (2013).
\newblock Women and minorities in computer science majors: Results on barriers from interviews and a survey.
\newblock {\em Issues in Information Systems}, 14(1):143--152.

\bibitem[Bouzguenda, 2013]{bouzguenda2013enablers}
Bouzguenda, K. (2013).
\newblock Enablers and inhibitors of learning transfer from theory to practice.
\newblock In {\em Transfer of learning in organizations}, pages 23--44. Springer.

\bibitem[Catet{\'e} et~al., 2020]{catete2020aligning}
Catet{\'e}, V., Alvarez, L., Isvik, A., Milliken, A., Hill, M., and Barnes, T. (2020).
\newblock Aligning theory and practice in teacher professional development for computer science.
\newblock In {\em Proceedings of the 20th Koli Calling International Conference on Computing Education Research}, pages 1--11.

\bibitem[Eckerdal, 2015]{eckerdal2015relating}
Eckerdal, A. (2015).
\newblock Relating theory and practice in laboratory work: A variation theoretical study.
\newblock {\em Studies in Higher Education}, 40(5):867--880.

\bibitem[Eckerdal et~al., 2024]{eckerdal2024learning}
Eckerdal, A., Berglund, A., and Thun{\'e}, M. (2024).
\newblock Learning programming practice and programming theory in the computer laboratory.
\newblock {\em European Journal of Engineering Education}, 49(2):330--347.

\bibitem[Gelonch-Bosch et~al., 2017]{gelonch2017teaching}
Gelonch-Bosch, A., Marojevic, V., and Gomez, I. (2017).
\newblock Teaching telecommunication standards: bridging the gap between theory and practice.
\newblock {\em IEEE Communications Magazine}, 55(5):145--153.

\bibitem[Giacaman and De~Ruvo, 2018]{giacaman2018bridging}
Giacaman, N. and De~Ruvo, G. (2018).
\newblock Bridging theory and practice in programming lectures with active classroom programmer.
\newblock {\em IEEE Transactions on Education}, 61(3):177--186.

\bibitem[Gretter et~al., 2019]{gretter2019equitable}
Gretter, S., Yadav, A., Sands, P., and Hambrusch, S. (2019).
\newblock Equitable learning environments in k-12 computing: Teachers’ views on barriers to diversity.
\newblock {\em ACM Transactions on Computing Education (TOCE)}, 19(3):1--16.

\bibitem[Hasker and Harriehausen-Muhlbauer, 2007]{hasker2007joint}
Hasker, R.~W. and Harriehausen-Muhlbauer, B. (2007).
\newblock A joint international master’s of computer science.
\newblock In {\em 2007 37th Annual Frontiers In Education Conference-Global Engineering: Knowledge Without Borders, Opportunities Without Passports}, pages T1A--7. IEEE.

\bibitem[Hayes and Overland, 2023]{hayes2023inclusive}
Hayes, L. and Overland, E. (2023).
\newblock {\em Inclusive Computing Education in the Secondary School: Linking Theory and Practice}.
\newblock Taylor \& Francis.

\bibitem[Janse~van Rensburg, 2020]{janse2020developing}
Janse~van Rensburg, J.~T. (2020).
\newblock {\em Developing guidelines for bridging the gap between IT theory and IT practice}.
\newblock PhD thesis, North-West University (South Africa).

\bibitem[{Jinshuju}, 2024]{jinshuju}
{Jinshuju} (2024).
\newblock {Jinshuju Online Survey Platform}.
\newblock Accessed: 19 October, 2024.

\bibitem[King, 2021]{king2021combining}
King, J. (2021).
\newblock Combining theory and practice in data structures \& algorithms course projects: An experience report.
\newblock In {\em Proceedings of the 52nd ACM Technical Symposium on Computer Science Education}, pages 959--965.

\bibitem[Kordaki and Berdousis, 2020]{kordaki2020identifying}
Kordaki, M. and Berdousis, I. (2020).
\newblock Identifying barriers for women participation in computer science.
\newblock {\em Pro Edu. International Journal of Educational Sciences}, 2(2):5--20.

\bibitem[Lessa and von Flach G.~Chavez, 2020]{FLOSS}
Lessa, M. S.~B. and von Flach G.~Chavez, C. (2020).
\newblock An approach for selecting floss projects for education.
\newblock In {\em Proceedings of the XXXIV Brazilian Symposium on Software Engineering}, SBES '20, page 463–472, New York, NY, USA. Association for Computing Machinery.

\bibitem[Martin, 2004]{martin2004addressing}
Martin, A. (2004).
\newblock Addressing the gap between theory and practice: It project design.
\newblock {\em Journal of Information Technology Theory and Application (JITTA)}, 6(2):5.

\bibitem[Murtaza et~al., 2024]{murtaza2024theory}
Murtaza, A., Fadare, S.~A., Mocsir, O.~M., Fadare, M.~C., Natividad, L.~R., Rafique, T., Akhtar, N., Shaheen, J., Mohsin, M., Taj, R., et~al. (2024).
\newblock From theory to practice: Harnessing ai for enhanced teaching-learning dynamics.
\newblock {\em Educational Administration: Theory and Practice}, 30(4):6331--6338.

\bibitem[Nascimento et~al., 2019]{nascimento2019does}
Nascimento, D. M.~C., von Flach Garcia~Chavez, C., and Bittencourt, R.~A. (2019).
\newblock Does floss in software engineering education narrow the theory-practice gap? a study grounded on students’ perception.
\newblock In {\em IFIP International Conference on Open Source Systems}, pages 153--164. Springer.

\bibitem[Pappas et~al., 2017]{pappas2017assessing}
Pappas, I.~O., Giannakos, M.~N., Jaccheri, L., and Sampson, D.~G. (2017).
\newblock Assessing student behavior in computer science education with an fsqca approach: The role of gains and barriers.
\newblock {\em ACM Transactions on Computing Education (TOCE)}, 17(2):1--23.

\bibitem[Phillips, 2005]{phillips2005challenging}
Phillips, R. (2005).
\newblock Challenging the primacy of lectures: The dissonance between theory and practice in university teaching.
\newblock {\em Journal of University Teaching and Learning Practice}, 2(1):2.

\bibitem[Randi and Corno, 2007]{randi2007theory}
Randi, J. and Corno, L. (2007).
\newblock Theory into practice: A matter of transfer.
\newblock {\em Theory into practice}, 46(4):334--342.

\bibitem[Reimer, 2019]{reimer2019computer}
Reimer, Y.~J. (2019).
\newblock Computer science in high school: Identifying and addressing common barriers.
\newblock {\em Journal of Computing Sciences in Colleges}, 35(1):14--21.

\bibitem[Resch and Schrittesser, 2023]{resch2023using}
Resch, K. and Schrittesser, I. (2023).
\newblock Using the service-learning approach to bridge the gap between theory and practice in teacher education.
\newblock {\em International Journal of Inclusive Education}, 27(10):1118--1132.

\bibitem[Samarasekara et~al., 2022]{samarasekara2022barriers}
Samarasekara, C.~K., Ott, C., and Robins, A. (2022).
\newblock Barriers to new zealand high school cs education-learners' perspectives.
\newblock In {\em Proceedings of the 53rd ACM Technical Symposium on Computer Science Education-Volume 1}, pages 927--933.

\bibitem[Samarasekara et~al., 2024]{samarasekara2024framework}
Samarasekara, C.~K., Ott, C., and Robins, A. (2024).
\newblock A framework identifying challenges \& solutions for high school computing.
\newblock {\em Education and Information Technologies}, pages 1--34.

\bibitem[Schulte and Knobelsdorf, 2007]{schulte2007attitudes}
Schulte, C. and Knobelsdorf, M. (2007).
\newblock Attitudes towards computer science-computing experiences as a starting point and barrier to computer science.
\newblock In {\em Proceedings of the third international workshop on Computing education research}, pages 27--38.

\bibitem[Scragg and Smith, 1998]{scragg1998study}
Scragg, G. and Smith, J. (1998).
\newblock A study of barriers to women in undergraduate computer science.
\newblock In {\em Proceedings of the twenty-ninth SIGCSE technical symposium on Computer Science Education}, pages 82--86.

\bibitem[Tedre and Pajunen, 2022]{tedre2022grand}
Tedre, M. and Pajunen, J. (2022).
\newblock Grand theories or design guidelines? perspectives on the role of theory in computing education research.
\newblock {\em ACM Transactions on Computing Education}, 23(1):1--20.

\bibitem[Thayer and Ko, 2017]{thayer2017barriers}
Thayer, K. and Ko, A.~J. (2017).
\newblock Barriers faced by coding bootcamp students.
\newblock In {\em Proceedings of the 2017 ACM Conference on International Computing Education Research}, pages 245--253.

\bibitem[Thayer, 2020]{thayer2020practical}
Thayer, K.~M. (2020).
\newblock {\em Practical Knowledge Barriers in Professional Programming}.
\newblock PhD thesis.

\bibitem[Wang and Hejazi~Moghadam, 2017]{wang2017diversity}
Wang, J. and Hejazi~Moghadam, S. (2017).
\newblock Diversity barriers in k-12 computer science education: Structural and social.
\newblock In {\em Proceedings of the 2017 ACM SIGCSE technical symposium on computer science education}, pages 615--620.

\bibitem[Y{\i}lmaz et~al., 2016]{yilmaz2016transition}
Y{\i}lmaz, R., {\c{S}}endurur, E., and {\c{S}}endurur, P. (2016).
\newblock Transition from theory to practice: a comparison of math and computer education and instructional technology teachers during school experience.
\newblock {\em Kastamonu Education Journal}, 24(3):1517--1532.

\bibitem[Young, 2003]{young2003bridging}
Young, L. (2003).
\newblock Bridging theory and practice: Developing guidelines to facilitate the design of computer-based learning environments.
\newblock {\em Canadian Journal of Learning and Technology/La revue canadienne de l’apprentissage et de la technologie}, 29(3).

\bibitem[Zyad, 2016]{zyad2016integrating}
Zyad, H. (2016).
\newblock Integrating computers in the classroom: Barriers and teachers' attitudes.
\newblock {\em International Journal of Instruction}, 9(1):65--78.

\end{thebibliography}

\section*{\uppercase{SURVEY QUESTIONNAIRE}}\label{sec:appendix}
Questions with standard bullets ($\bullet$) are single-select while questions
with square ($\square$) are multi-select.

1) What is your age? $\bullet $ Under 18  $\bullet $ 18-23  $\bullet $ 24-29  $\bullet $ 30 and above  

2) What is your gender? $\bullet $ Male $\bullet $ Female $\bullet $ Non-binary $\bullet $ Prefer to self-describe [Text field] $\bullet $ Prefer not to say 

3) What is your ethnic group? $\bullet $ Han $\bullet $ Other [Text field] $\bullet $ Prefer not to say

4) What subject are you studying in the last semester? $\bullet $ Computer Architecture $\bullet $ Electronics and IoT $\bullet $ Automation and Embedded Systems $\bullet $ Other [Text field]

5) What year of study are you in? $\bullet $ First Year $\bullet $ Second Year $\bullet $ Third Year $\bullet $ Fourth Year 

6) How would you rate the difficulty of understanding theoretical concepts in your computing courses?
$\bullet $ Very Easy $\bullet $ Easy $\bullet $ Moderate $\bullet $ Difficult $\bullet $ Very Difficult

7) How challenging do you find applying theoretical knowledge to practical tasks in your computing courses? 
$\bullet $ Very Easy $\bullet $ Easy $\bullet $ Moderate $\bullet $ Difficult $\bullet $ Very Difficult 

8) Please select exactly three barriers you believe have the greatest impact on your computing courses. If your answers are not among the options, please add them below.

$\square$ Language barriers in understanding course material

$\square$ Insufficient prior knowledge of key computing concepts

$\square$ Difficulty in grasping abstract theoretical concepts

$\square$ Challenges due to cultural differences in learning environments

$\square$ Unclear explanations of the knowledge from lecturers 

$\square$ Lack of practical examples in the course material

$\square$ Inadequate lab facilities or hands-on practice opportunities 

$\square$ Limited access to necessary technology or software

$\square$ The inappropriate assessment of the course (e.g. the weights of theory and practice are unreasonable.) 

$\square$ The improper course content composition, such as having much theory and not much practical content 

$\square$ Peer collaboration or group work challenges

$\square$ Other [Text field]

9) To what extent does the chosen barrier 1 affect your learning? 
$\bullet $ Not at all $\bullet $ Slightly $\bullet $ Moderately $\bullet $ Very $\bullet $ Extremely

10) To what extent does the chosen barrier 2 affect your learning? 
$\bullet $ Not at all $\bullet $ Slightly $\bullet $ Moderately $\bullet $ Very $\bullet $ Extremely

11) To what extent does the chosen barrier 3 affect your learning? 
$\bullet $ Not at all $\bullet $ Slightly $\bullet $ Moderately $\bullet $ Very $\bullet $ Extremely

12) What was your level of experience in computing before entering the current course?
$\bullet $ No Experience $\bullet $ Beginner $\bullet $ Intermediate $\bullet $ Expert

[If you selected 'Beginner,' 'Intermediate,' or 'Expert,' please answer Question 13A).:]

13A) To what extent do you think your previous computing experience has helped you perform well in these courses?
$\bullet $ Not at all $\bullet $ Slightly $\bullet $ Moderately $\bullet $ Very $\bullet $ Extremely

14) How engaged are you in your computing courses?
$\bullet $ Not at all $\bullet $ Slightly $\bullet $ Moderately $\bullet $ Very $\bullet $ Extremely 

[If you selected any option other than 'Not at all', please answer Question 15A):]

15A) Do you think your level of participation in the course affects your learning and helps you overcome learning barriers? 
$\bullet $ Strongly disagree $\bullet $ Disagree $\bullet $ Neutral $\bullet $ Agree $\bullet $ Strongly agree

16) To what extent has the support you received from the lecturer and teaching assistant helped you overcome obstacles?
$\bullet $ Not at all $\bullet $ Slightly $\bullet $ Moderately $\bullet $ Very $\bullet $ Extremely

17) How has transitioning from the Chinese educational system to the New Zealand educational system affected your learning experience and outcomes?
$\bullet $ Negative $\bullet $ Slightly negative $\bullet $ Neutral $\bullet $ Slightly positive $\bullet $ Positive

18) What aspects do you think can be further improved in these courses that have both theoretical and practical components? Please provide as much detail as possible. [Open-ended question. Participants were allowed to write without any word limit.]

19) Do your interests and career aspirations fall into the computing/computer science field?
$\bullet $ Strongly disagree $\bullet $ Disagree $\bullet $ Neutral $\bullet $ Agree $\bullet $ Strongly agree

[If you selected 'Strongly disagree' or 'Disagree' in Question 19), please answer Question 20A):]

20A) Do your parents, relatives, and friends influence your career aspirations?
$\bullet $ Strongly disagree $\bullet $ Disagree $\bullet $ Neutral $\bullet $ Agree $\bullet $ Strongly agree

[If you selected 'Neutral,' 'Agree,' or 'Strongly agree' in Question 19), please answer Question 20B):]

20B) Do your career aspirations align with your interests?
$\bullet $ Strongly disagree $\bullet $ Disagree $\bullet $ Neutral $\bullet $ Agree $\bullet $ Strongly agree

[If you answered Question 20A), please answer Question 21A):]

21A) Because your future career is unrelated to computer technology, do you find it more difficult to participate in this course?
$\bullet $ Strongly disagree $\bullet $ Disagree $\bullet $ Neutral $\bullet $ Agree $\bullet $ Strongly agree

[If you answered Question 20B), please answer Question 21B):]

21B) Does your education in the computing field help motivate you and make it easier to overcome barriers encountered during the course?
$\bullet $ Strongly disagree $\bullet $ Disagree $\bullet $ Neutral $\bullet $ Agree $\bullet $ Strongly agree

22) For courses like this, do you have any specific study strategies or recommendations on learning resources for your juniors? Please provide as much detail as possible. [Open-ended question. Participants were allowed to write without any word limit.]

\begin{comment}
\begin{table*}[h!]
\centering
\caption{Mapping Barriers, Effects, and Proposed Solutions}
\begin{tabular}{|p{4cm}|p{5cm}|p{5cm}|}
\hline
\textbf{Barrier} & \textbf{Effect on Students} & \textbf{Proposed Solution} \\
\hline
Insufficient prior knowledge & Difficulty understanding theoretical concepts and applying them & Preparatory courses, mentorship, and technical vocabulary courses \\
\hline
Language barriers & Reduced comprehension and increased academic stress & Bilingual materials, cultural integration workshops, and English Corners \\
\hline
Differences in educational systems & Challenges adapting to new learning methods & Orientation sessions and linked cultural awareness workshops \\
\hline
Limited access to technology & Inability to complete practical tasks effectively & Computer labs, loaner laptops, and technical training sessions \\
\hline
Overwhelming workload & High stress and reduced performance & Balanced curricula and time management workshops \\
\hline
Lack of interest in computing & Low engagement and unclear career pathways & Career counselling, mentorship with industry professionals, and highlighting real-world applications \\
\hline
\end{tabular}
\end{table*}
\end{comment}

\end{document}